\documentclass[11pt]{article}

\usepackage{amssymb,amsmath}
\usepackage{amsfonts}
\usepackage{color,float}
\usepackage{ graphicx, latexsym, amssymb, amscd, psfrag, eepic}

\newcommand{\smallgap}{\vspace{0.25in} \noindent}
\newcommand{\be}{\begin{equation}}
\newcommand{\ee}{\end{equation}}

\begin{document}

\long\def\comment#1{}

\newcommand{\ind}{ {\rm I\kern-0.19emI} }
\newcommand{\sign}{\mbox{\rm Sign}}
\newcommand{\prob}{I\kern-0.37emP}
\newcommand{\leg}{\
\begin{array}{ccc}
  > \\[-.08in]
  = \\[-.08in]
  <
\end{array}
\ }
\newtheorem{theo}{Theorem}
\newtheorem{lem}{Lemma}
\newtheorem{cor}{Corollary}
\newtheorem{prop}{Proposition}
\newcommand{\Z}{{\mathbb Z}}
\newcommand{\E}{{\mathbb E}}
\newcommand{\N}{{\mathbb N}}
\newcommand{\R}{{\mathbb R}}
 \newenvironment{remark}{\noindent\textbf{Remark. }}{}

\newenvironment{acknowledgement}{\bigskip\noindent\textbf{Acknowledgement:}\\}{}
\def\Def{{\bf {\textsf Definition.}}~}
\def\Ex{{\bf Example.}~}
\def\proof{{\it Proof.}~}

\title{Alarm System for Insurance Companies: \\ A Strategy for Capital Allocation }
\date{}

\author {S. Das \thanks {IIM Bangalore, India} \qquad and \qquad M. Kratz \thanks{ESSEC Business School Paris, France} ~\thanks {MAP5 (UMR8145, Univ. Paris Descartes)}}
\maketitle


\begin{abstract}
\noindent One possible way of risk management for an insurance company is to develop an early and appropriate alarm system before the possible ruin. The ruin is defined through the status of the aggregate risk process, which in turn is determined by premium accumulation as well as claim settlement outgo for the insurance company. The main purpose of this work is to design an effective alarm system, i.e.\ to define alarm times and to recommend augmentation of capital of suitable magnitude at those points to prevent or reduce the chance of ruin.
To draw a fair measure of effectiveness of alarm system, comparison is drawn between an alarm system,  with capital being added at the sound of every alarm, and the corresponding system without any alarm, but an equivalently higher initial capital. Analytical results are obtained in general setup and this is backed up by simulated performances with various types of loss severity  distributions. This provides a strategy for suitably spreading out the capital and yet addressing survivability concerns at satisfactory level.
\end{abstract}

\smallgap {\it AMS classification.} 91B30, 60K30

\section{Introduction and Overview}

This work develops an early and appropriate alarm system for an insurance institution before its possible ruin based on pattern of premium collection and demands for claim settlement. While keeping a very high initial capital may avoid ruin for the insurance company, it is neither desired by most companies because of obvious investment concerns, nor is feasible at times. An effective alarm system opens the door for an alternate strategy based on ruin theory by opting for less initial capital and topping it up when really necessary.

The work may be applied from two perspectives. On one hand, it can be viewed from a regulatory perspective to provide guidance for regulatory intervention, without compromising the capacity of a company to survive. On the other hand, companies may use it to design good triggers for obtaining contingent capital from banks. The alarm system could serve in structuring such contract to improve capital management.

Alarm systems have been developed in different contexts in the literature (viz. \cite{Lindgren}, \cite{ZD2O}, \cite{Scotto}, \cite{GKL},  and references therein), while capital reserving or capital allocation have been addressed in many articles (viz.\ \cite{BDMKM}, \cite{KDI}, and references therein). In particular, in \cite{KDI} Kaishev et al.\ showed numerically  that two capital accumulation functions, one linear and the other piecewise linear with one jump at some instances, would lead to equal chances of survival and also equal accumulate risk capital at the end of the considered time interval. The approach in the present work, with the introduction of a new alarm system,  is fundamentally different even though the broad concern is similar, i.e.\ to reduce the initial capital without compromising on the survival probability.

\noindent The basic idea behind our proposed notion of alarm is as follows.  Alarm is sounded at a juncture when the probability of ruin (in absence of any intervention) within a specified future time period is high enough. While  few variations  in defining the alarm time have been explored in \cite{wpDK}, we find it more appropriate when the above probability is set in terms of conditional event given survival up to the alarm time. In addition, we require that the chance of no ruin before the alarm should be sufficiently high.  The natural extension of single alarm when adding capital at the sound of each alarm leads to the definition of an alarm system consisting of successive alarms. This system constitutes an alternate strategy for having to put up an excessive initial capital to avoid ruin.

Note that this strategy does not interfere with the Value-at-Risk approach (or any tail approach) applied by insurance companies as mandated by the Solvency regulation. It just means that the capital may also be adjusted on a regular basis (e.g.\ every quarter) for the risk adjusted capital to be higher than the capital required by Solvency.

For fair evaluation of effectiveness of our strategy, the proposed alarm system is pitted against a default no-alarm system equipped with equivalent higher initial capital. We compare the survival probabilities under the alternatives. In the longer run, the alarm system is expected to win, as is indeed confirmed by our study. In the shorter run, the alarm system may be preferred even if the chance of survival under this is marginally worse. With that being the objective, we focus on analytical as well as numerical evaluation of the comparative survival probabilities under the two systems.

To illustrate our method, we consider a simple linear accumulation model. So, the adjustment of capital as mandated by Solvency rules might be easily accommodated in the setup using a stepwise linear accumulation function. Moreover, this simple framework is not a prerequisite for the proposed formulation and essence of our work. It would be valid irrespective of whether the claim amounts are independent, or identically distributed, or otherwise. Thus our method has the advantage of being simple and adaptable to any model. Again, for the sake of simplicity, in particular in the numerical setup, we have stuck to i.i.d.\ claims.
If the stochastic nature of the risk process is completely known, as assumed in this work, the alarm times are naturally fixed known parameters, depending on various parameters of the underlying risk process.
In practice, the proposed mechanism may be embedded into an adaptive scheme where additional information regarding the risk process in terms of claims would be recursively/progressively utilized to lead to a suitable random alarm system that draws on empirical information on claims.

\noindent The paper is organized as follows. In Section~{\ref{framework}, we introduce the basic notation and framework of the work, as well as key results on ruin time distributions from the existing literature. The formal definition of alarm time is given in Section~\ref{definition}.
In Section~\ref{numerical}  we choose few examples in simulation settings to cover different types of severity distributions, discrete to continuous, as well as light vs.\ heavy tail, and study the role of various parameters in the alarm times. Formalization of multiple alarms leading to alarm system is taken up in Section~\ref{multiple:alarm}.
The next section, Section~\ref{alarm:compare}, develops a strategy to alleviate initial capital using alarm systems.
The  effectiveness of alarm systems and comparison across the different options including that of not adopting any alarm system is discussed here. The  numerical demonstrations are provided in Section~\ref{compare:numerical}. General analytical bounds are derived in Section~\ref{compare:analytical}, providing directions of adaptability of the alarm system in specific real circumstances.

\section{Alarm System Based on Probability of Impending Ruin} \label{sec:alarm2}

\subsection{Framework}\label{framework}

\subsubsection{Notation} \label{notation}

We consider the classical ruin theory model, the Cram\'er Lundberg model, to present our general approach.
To keep the framework as general as possible, most of the definitions and results will be given in terms of the ruin time probability, the results on (joint) c.d.f.\ of ruin time for dependent or independent claims being found in the existing literature.
However,  for  comparing our formal results with simulations,  we consider simpler setup with i.i.d.\ simulated claims, since the aim is   only to illustrate our method, although it may be extended for dependent claims using e.g.\ copula methods.
Much of the analysis of this paper can be carried over in a straightforward way to more general L\'evy processes and premium rates.

\noindent Let us assume that the $i$-th claim is of magnitude (severity) $X_i$ and happens at time $T_i$, for $i \ge 1$. By default, we assume that $X_i$'s are with distribution function $F$ and  mean $\mu$, with i.i.d.\ inter-arrival times $T_k-T_{k-1}$ also independent of $(X_i)$; a common model for the claim process is to be a Poisson process. For most part we consider such Poisson process but with various different claim distributions $F$; however most of the general formulations and definitions described in this work may be valid under more general framework.

\noindent Set $T_0=0$. Let $(N_t)_{t\ge 0}$ defined by $N_t=sup\{k\ge 1: T_k\le t\}$ be a Poisson process with intensity $\lambda>0$, independent of the $(X_i)$.
Aggregated claims $(S_t)_{t\ge 1}$ are defined by $S_t=\sum_{i=1}^{N_t}X_i$.

\smallgap Consider the risk (or surplus) process $(V_u(t))_{t\ge 0}$  given by
\be
V_t= u_t + p_t - S_t=u_t - R_t,
\ee
where $u_t$ denotes the capital function at time $t$ and the premium rate is linear, viz.\ $ p_t = ct$ and the net outgo (without taking capital into account, i.e.\ aggregate claims less premium collected) process is given by
$R_t= S_t- p_t $. Note that while $R_t$ is a stochastic process (a L\'evy process, more precisely a compound Poisson process in our framework), the capital process $ u_t$ is non-random and at the discretion of the company. Indeed, one of the key objective of this work may restated as determination of $u_t$ given the knowledge of parameters of $R_t$.

If the decision is to start with only an initial capital $u_0=u$ and not make any further addition, then $u_t = u$ for all $t \ge 0$; in such a case, we may denote the risk process, equivalently by $V_u(t)$. This is indeed the benchmark or starting framework.
The ruin time
of such a risk process is then formally defined as:
\be
T(u)=\inf\{t>0: V_t <0\}=\inf\{t>0: R_t>u\},
\ee
with $T(u)=\infty$ if there is no ruin. Note that while in practice one may wish to define ruin as the first time instance when $V_t$ goes below a level $L$ (other than 0),  it would take a trivial adjustment in the approach adopted here to carry forward the method. Consequently, in this work, we stick to $L=0$.

\smallgap Recall that for the classical model, the Net Profit Condition (NPC) is given by: $c>\lambda\mu$, {\it i.e.} the premium income should exceed expected claim payments.
Introducing the premium loading factor $\theta$, we can write $c$ as
\be
c=(1+\theta)\lambda\mu, \label{NPC}
\ee
so the NPC is equivalent to $\theta>0$.\\

\begin{remark}
If $\theta>0$ is small (large), it reflects heavy (light) traffic condition.
It is, of course,  safer for the company to have NPC satisfied;  but, in practice it may not always be possible. An important group of examples of such situations are heavy-tail severity distribution having infinite mean; naturally NPC is violated in such conditions with $\theta$ being -1.
Consequently we strive for formulation and general results that do not depend on the NPC and in some of the computational experiments, NPC is violated.
\end{remark}

\begin{remark}
Note that while flexibility in the choice of the initial capital $u$ is an integral part of this work, in a given instance we are concerned with a fixed value for $u$ and {\em not} in the asymptotic behaviour with $u \rightarrow \infty$, unlike most related literature. The justification for such asymptotic consideration (\cite{Mikosch}) is usually given in terms of avoid being in a framework of infinite horizon ruin probability of one. However, not only a fixed initial capital is practical, it is perhaps inevitable (and hence acceptable) that any system  will {\it eventually } ruin
if not interrupted/not intervened at some finite time. It is also from this consideration that we do not insist on NPC in the present work under all circumstances.
\end{remark}

\smallgap Let $T(a,u)$ denote the (first) ruin time after $a$ of the risk process $(V_u(t); t\ge a)$ with initial capital $u$; i.e.\ 
$$
T(a,u)=\inf\{t > a: V_u(t)<0\}=\inf\{t >  a: R_t>u\}.
$$
Set $ T(u) \equiv T(0,u)$.
The infinite horizon ruin probability with capital $u$ at time $a$ is denoted by:
$$
\psi_a(u):=P[T(a,u)<\infty]=P[\inf_{t > a}V_u(t)<0]=P[\sup_{t > a}R_t>u],
$$
and the corresponding finite horizon ruin probability, which is the distribution function of the r.v. $T(a,u)$, by
$$
\psi_a(u,t):=P(T(a,u)\le t])=P[\sup_{a< s\le t} R_s >u];
$$
to simplify the notation, we set
$$
\psi_0(u)=\psi(u)\quad \mbox{and}\quad \psi_0(u,t)=\psi(u,t)$$
and also $$\bar\psi_a(u,t)=1-\psi_a(u,t).$$

\noindent Let us introduce the conditional ruin probabilities in infinite and finite times given some event $B$:
$$
\psi_a(u~|~B):=P[T(a,u)<\infty~|~B]=P[\inf_{t > a}V_t<0~|~B]=P[\sup_{t > a}R_t>u~|~B],
$$
and
$$
\psi_a(u,t~|~B):=P(T(a,u)\in (0,t]~|~B)=P[\sup_{a\le s\le t}R_s >u~|~B].
$$
For $B$ with $P[B=0]$, as would be the case if $B=(Z = z)$ for any continuous random variable $Z$, the above conditional probabilities should be interpreted (understand) in usual manner, namely as conditional expectations:
$$\psi_a(u,t~|~Z(\omega)=z)=E[1_{\big(\sup_{a\le s\le t}R_s >u\big)}~|~Z](\omega);$$
nevertheless to alleviate the notation, we will keep the notation of conditional probability even in this case and all over the paper.\\

\noindent Notice that, $a$ being fixed, $\psi_a(u,t)$ is a non-decreasing function of the time $t$ at given $u$, and a non-increasing function of the capital $u$ at given $t$. \\

\smallgap Finally, introducing a given time $a>0$ and the notation $F_{R_a}$ for the cumulative distribution function of $R_a$, we can write
$$
\bar{\psi}(u) = P[\sup_{t > 0} R_t \le u]=\int_{-\infty}^u P[\sup_{t > a} R_{t-a} \le u-x]~P[\sup_{0 < t \le a } R_t  \le u ~|~ R_a=x]~dF_{R_a}(x)
$$
since   $(R_t)$ have independent and stationary increments (in particular, $R_t-R_a $ is independent of $\sigma \{ R_s: 0 < s \leq a \}$ and $R_t-R_a \overset{d}{=}R_{t-a}$),
\begin{equation}\label{indpdceInfinite}
\mbox{\it i.e.}\qquad\bar{\psi}(u) ~=~\int_{-\infty}^u \bar{\psi}(u-x) \bar{\psi}\big(u,a~|~ R_a=x \big)~dF_{R_a}(x).
\end{equation}

\noindent A lower bound for $\bar{\psi}(u)$ can then be deduced, using that $\psi(u)$ is a non-increasing function of $u$, as
$$
\bar{\psi}(u) ~\ge ~\bar{\psi}(0)~\int_{-\infty}^u \bar{\psi}\big(u,a~|~ R_a=x \big) ~dF_{R_a}(x) = \bar{\psi}(0)~\bar{\psi}\big(u,a\big).
$$

\smallgap Looking at finite time ruin probabilities, we can proceed in the same way and obtain for $t>a$ (the alternate case being trivial)
\begin{equation}\label{indpdceFinite}
\bar{\psi}(u,t) ~=~ P[\sup_{0<s \le t} R_s  \le u]
= \int_{-\infty}^u \bar{\psi}(u-x,~t-a)~\bar{\psi}\big(u,a~|~ R_a=x \big)~dF_{R_a}(x)
\end{equation}
and we have
$$
\bar{\psi}(u,t) ~\ge ~\bar{\psi}(0,t-a)~\bar{\psi}\big(u,a\big).
$$

\smallgap Notice that the infinite time ruin probabilities can be deduced from the finite ones when taking $t\to\infty$.

\smallgap Finally, introducing the $\sigma$-field $\sigma_a:=\sigma\{R_s; ~s\le a\}$, let us write the last useful relation which holds for any times $0<a<t ~(\le\infty)$ and for any $u$ and almost all $x$,
\begin{eqnarray}\label{psiconditio}
\psi_a\big(u,t~|~\sigma_a~\cap ~(R_a=x)\big)
&=& P[\sup_{a<s\le t}R_s -R_a >u-x ~|~\sigma_a~\cap ~(R_a=x)]\nonumber\\
(\mbox{by independence and stationarity})&=& P[\sup_{a<s\le t}R_{s-a} >u-x]~=~ \psi\big(u-x~,t-a\big).
\end{eqnarray}
In particular, for $t\to\infty$, ~$\displaystyle \psi_a(u~|~~\sigma_a~\cap ~(R_a=x))=\psi\big(u-x\big)$.

\subsubsection{Key relevant results from the existing Literature} \label{literature}

Distributions of ruin time have been explored extensively in the literature, \cite{Asmussen}, \cite{Dickson}, \cite{EKM}, \cite{IK04}, \cite{IK06}, \cite{KD}, \cite{Mikosch} being some key references in this decade. The alarm times we will define in this paper are simply a parameter of the ruin time distribution, therefore we will attempt to calculate them using such characterizations of ruin-time. From broad considerations,
ruin time distributions depend on whether the claims distributions are continuous or discrete, but also on the characterization of the claims tail distributions. Our choice of examples reflect that in attempt to cover broad perspectives.\\

\noindent Let us recall, for completeness and to help the reading, some exact evaluations of ruin probabilities that can be found in the literature.

\smallgap $\bullet$ {\small\bf\textsf Exponential i.i.d.\ claims}\\
It is one of the few cases where ruin probabilities can be explicitly computed.\\
When considering a light-tailed (i.i.d.) claim distributions in the form of exponential distribution with parameter $\rho>0$, the ultimate ruin probability is given, under the NPC, by
\begin{equation}\label{ultimateRPexp}
\psi(u)=\frac1{1+\theta} ~e^{-\frac{\rho~\theta}{1+\theta}~u},
\end{equation}
with $\theta$ the loading premium factor,
which corresponds to the Lundberg's bound up to the multiple factor $(1+\theta)^{-1}$. \\
The ruin time c.d.f.\ $\psi(u,.)$ can also be explicitely/analytically computed for exponential claims (see e.g. \S IV in \cite{Asmussen}, or \S 8 in \cite{Dickson} when expressing the Bessel function as a series instead of an integral) according to
\begin{equation}\label{finiteRPexp}
\psi(u,t)=\psi(u)-\frac1\pi \int_0^{\pi}\frac{f_{u}(t,x)g_{u}(x)}{h(x)}~dx,
\end{equation}
where
\begin{eqnarray*}
f_{u}(t,x)=f_{\rho,\theta,u}(t,x)&=& \frac1{1+\theta}\exp\left\{\frac{\rho~(u+2~t)}{\sqrt{1+\theta}}~\cos x-\rho\left(u+\frac{2+\theta}{1+\theta}~t\right)\right\},\\
g_u(x)=g_{\rho,\theta,u}(x)&=& \cos\left(\frac{\rho~u}{\sqrt{1+\theta}}~\sin x \right)-\cos\left( \frac{\rho~u}{\sqrt{1+\theta}}~\sin x ~+~ 2x\right),\\
h(x)=h_{\theta}(x)&=& \frac{2+\theta}{1+\theta}-\frac{2}{\sqrt{1+\theta}}~\cos x.
\end{eqnarray*}

\smallgap $\bullet$ {\small\bf\textsf Exact evaluation of finite time survival probability for any claim} (Ignatov et al., \cite{IK04}, \cite{IKK01})\\

\noindent If $f_X$ denotes the joint density function
for {\it continuous claims} $X$, the finite time survival probability is given by (see \cite{IK04}, Theorem 1)
\begin{eqnarray}\label{contFRTproba}
\bar{\psi}(u,t) &=&\!\!\!e^{-\lambda t}~(~1+\sum_{k=1}^{\infty}\lambda^k \int_0^{u+p(t)}\!\!\!\!\!\!dy_1\int_{y_1}^{u+p(t)}\!\!\!\!\!\!dy_2 \cdots \int_{y_{k-1}}^{u+p(t)} \!\!\!\!\!A_k\left(t;\nu_{y_1},...,\nu_{y_k}\right)\nonumber\\
&& \qquad\qquad \qquad\qquad\qquad \quad f_X(y_1, y_2-y_1,...,y_k-y_{k-1})~dy_k ~)
\end{eqnarray}
where
$\displaystyle A_k(t;\kappa_1,...,\kappa_k)$, $k\ge 1$, are the classical Appell polynomials $A_k(t)$ of degree $k$ with a coefficient in front of $t^k$ equal to $1/k!$, defined for $k\ge 1$ by
$\displaystyle A_0(t)=1$, $\displaystyle A_k'(t)=A_{k-1}(t)$ and $\displaystyle A_k(\kappa_k)=0$, and for $i\ge 0$, $\nu_i=\inf\{t: u+p(t)\ge i\}$.\\~\\
For {\it discrete claims}, it is given by (see \cite{IKK01})
\begin{eqnarray}\label{discretFRTproba}
\bar{\psi}(u,t)
&=& e^{-\lambda t}\sum_{k=1}^n\!\!\!\!\!\sum_{
{\tiny\begin{array}{c}
\sum_{i=1}^{k-1} x_i\le n-1\\
x_i\ge 1,1\le i\le k-1
\end{array}}
}
\!\!\!\!\!\!\!\!\!\!\!\!\!P\left[X_i=x_i, 1\le i\le k-1; X_k\ge n-\sum_{i=1}^{k-1}x_i\right]\nonumber\\
&& \qquad\qquad\qquad \qquad\quad\sum_{j=0}^{k-1}(-1)^jb_j(z_1,...z_j)\lambda^j\sum_{m=1}^{k-j-1}(\lambda t)^m/m!
\end{eqnarray}
where $n$ is the integer part of $(1+u+p(t))$, and for $j\ge 1$, $z_j=\nu_{{\tiny \sum_{i=1}^j x_i}}$,\\
$\displaystyle b_j(z_1,...z_j)= \sum_{i=0}^{j-1} \frac{(-1)^{j+i+1}}{(j-i)!}z_j^{j-i}b_i(z_1,...,z_i)$, with $b_0(.)=1$ and $b_1$ the identity function.

\subsection{Defining Alarm Time} \label{definition}

\noindent Different approaches have been proposed so far in the literature, defining random or deterministic alarm times; see e.g. \cite{Lindgren}, \cite{Scotto} and references therein, \cite{GKL}, \cite{wpDK}).

In this paper, we propose a new approach for  devising deterministic alarms based on two intuitive requirements of the alarm time,  viz.\
 \begin{itemize}
   \item at the alarm time, the chance of ruin in {\em not so distant future} is {\em substantial} if no remedial action is taken;
   \item the chance of the system getting ruined before this (alarm) time is {\em minimal}.
 \end{itemize}
 The formal definition, included below, puts these requirements objectively.

\smallgap
\Def Given specified probabilities $\alpha$ and $\beta$  and future (lead) time window $d$ (to be chosen by the company), we define the alarm time $A =A (\alpha, \beta,d,u)$ as:
\begin{eqnarray}\label{def_A}
A&=&\inf \left\{ s > 0:  P[ T(u) \le s+d ~|~ T(u) > s] \geq 1 - \alpha ~\mbox{and}~  P[ T(u) > s] \geq 1 -\beta
\right\}.
\end{eqnarray}
~\vspace{0.05cm}

\noindent With $\psi(u,.)$ denoting the c.d.f.\ of ruin time $T(u)$ with initial capital $u$, we have
$$
P[T(u)\le s+d~|~T(u)>s] \ge 1 - \alpha \Leftrightarrow  \alpha~\bar{\psi}(u,s)~-~\bar{\psi}(u,s+d)\ge 0 $$
Hence to identify  the alarm time $A$ , for given $d$ and $\alpha$, we look for the first time
$s>0$ satisfying
$$ \alpha~\bar{\psi}(u,s) ~-~\bar{\psi}(u,s+d)\ge 0, \quad \mbox{\rm and } \quad \bar{\psi}(u,s)\ge 1-\beta, $$
\begin{equation}\label{equiv:alarm2}
    \mbox{or equivalently } A = \inf_{s>0} \Big\{ s~:~
\bar{\psi}(u,s)~\ge~ \max\big(1-\beta~,~\frac1\alpha~\bar{\psi}(u,s+d) \big) \Big\}.
\end{equation}

{\bf Choice of $\alpha$, $\beta$ and $d$}.} The parameter $\beta$ in the above specification requirement need to be  small ensuring that the chance of system getting ruined before the sound of alarm is insignificant. On the other hand, $ \alpha$ should be (only) moderately small to ensure that the threat of ruin is realistic enough to warrant a remedial action. To emphasize, if $\alpha = 0.4$, there is a 60\% chance of ruin in the given future window, which is bad enough for one to consider options available and one need not (actually should not) have to wait for a situation where this chance is very close to 1 (which would be the case, if we were to demand say $\alpha =0.01$).  The time window $d$ should be moderate, since  a `large' value would imply that the  ruin is far from being imminent (and hence perhaps the threat is not very serious in that sense), while a small value would give very little opportunity for the remedial actions to take any effect. While in practical situations these choices would be somewhat subjective and/or depending on other problem specific elements, the choice of $\alpha$ and $d$ would be inter-related.
We also note that the alarm time, so defined, is actually non-random.
We may also note that,  a simple consequence of our definition of alarm time $A$ is:
$$ P[ A < T(u) < A + d] \ge (1-\alpha)(1-\beta).$$

\subsection{Numerical illustrations via simulation} \label{numerical}

\noindent We undertake simulation study to better understand the alarm times defined in (\ref{def_A}), specially in terms of  how it varies with the parameters $\alpha$, $\beta$ and $d$, but also how it depends on the initial capital.
To cover discrete and continuous distributions with different tails, heavy vs.\ light,  for claim severities (density $f(\cdot)$), we take three examples in the form of  Exponential, Pareto and (discrete) Logarithm distributions. In all these examples, occurrence of claims are assumed to constitute Poisson process, i.e.\ the time between successive claims follows an Exponential distribution ($\lambda$); also the risk process entails an initial capital $u_0$ and a linear premium function $p_t = c t$. All results are based on simulation run of 100000 carried in R. For the sake of brevity, only  selected tables and figures are reported given below; for additional results see \cite{wpDK}.


\begin{figure}[h] 
\begin{center}
{\bf \scriptsize Figure 1: Density of Ruin-time}

$\begin{array}{cc}
\includegraphics[width=3in]{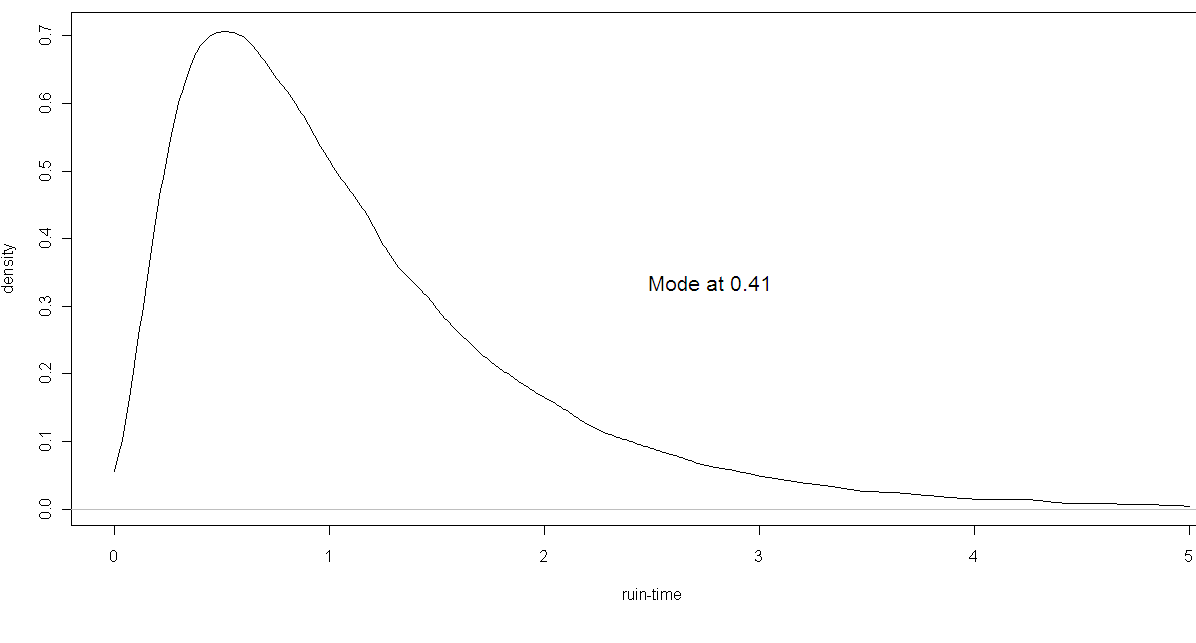} &
\includegraphics[width=3in]{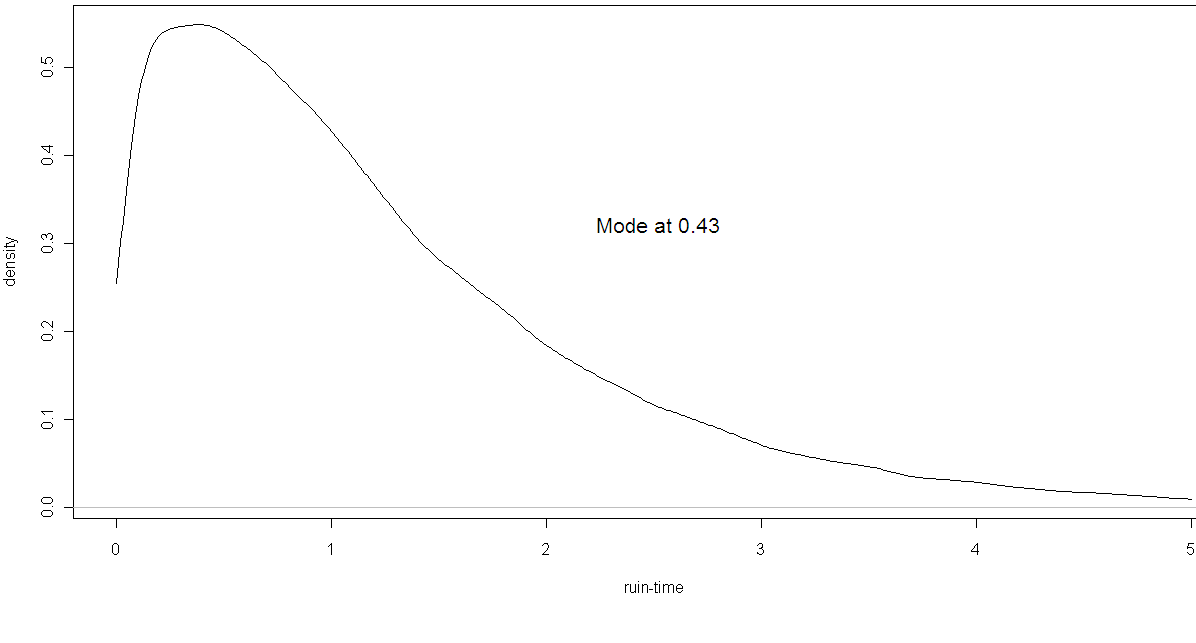} \\
{\rm Example \; 1: Exponential \; severity} ,   &
{\rm Example \; 2: Pareto \; severity}    \\
f(x)= \rho \exp(-\rho x), \quad x>0 \quad (\rho=0.5) &
f(x)= \frac{\kappa \rho^{\kappa}}{(\rho+x)^{\kappa+1}} \quad x>0 \quad (\rho=1 , \kappa=0.95) \\
u_0= 15, c = 25, \lambda= 20, & u_0= 50,  c = 40,  \lambda= 20,\\
\end{array}$
\includegraphics[width=3in]{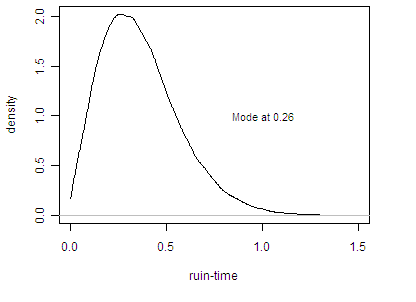} \\
{\rm Example \; 3: Logarithm severity} \\
$f(i)= \frac{- \kappa^i }{i \times \ln (1- \kappa)}, \quad i=1,2, \ldots \quad  (\kappa=0.95$) \\  $u_0= 50,  c = 25,  \lambda= 30$
\end{center}
\end{figure}

\subsubsection{How alarm time varies with its various parameters}

\noindent We compute the alarm time with the various choices of $\alpha, \beta$ and $d$. Tables 1.1a and Tables 1.1b are respectively for $\beta=0.025$ and 0.25 in the context of Example 1, while Tables 1.2 and 1.3 for Examples 2 and 3, for various combination of $d$ and $\alpha$. No alarm is noted as NA in the tables below and instantaneous alarm as 0.

\begin{center}
{\bf \scriptsize Table 1.1a:  Alarm times with $\beta =0.025$, for different $d$ and $\alpha$ -- Exponential case (Ex.1)}
{\scriptsize
\begin{tabular}{|c|ccccccccc|} \hline
         d  &                                                                                                                  \multicolumn{ 9}{|c|}{$\alpha$} \\
 &  {\bf 0.3} & {\bf 0.325} & {\bf 0.35} & {\bf 0.375} &  {\bf 0.4} & {\bf 0.425} & {\bf 0.45} & {\bf 0.475} &  {\bf 0.5} \\
\hline
      0.75 &               NA &         NA &         NA &         NA &         NA &         NA &         NA &         NA &         NA \\

       0.8 &               NA &         NA &         NA &         NA &         NA &         NA &         NA &         NA &       0.12 \\

      0.85 &               NA &         NA &         NA &         NA &         NA &         NA &         NA &       0.11 &       0.06 \\

       0.9 &               NA &         NA &         NA &         NA &         NA &         NA &       0.11 &       0.05 &          0 \\

      0.95 &               NA &         NA &         NA &         NA &         NA &        0.1 &       0.05 &          0 &          0 \\

         1 &               NA &         NA &         NA &         NA &       0.11 &       0.05 &          0 &          0 &          0 \\

      1.05 &               NA &         NA &         NA &       0.11 &       0.05 &          0 &          0 &          0 &          0 \\

       1.1 &               NA &         NA &       0.12 &       0.06 &          0 &          0 &          0 &          0 &          0 \\

      1.15 &               NA &       0.14 &       0.06 &          0 &          0 &          0 &          0 &          0 &          0 \\
\hline
\end{tabular}
}
\end{center}

\begin{center}
{\bf \scriptsize Table 1.1b:  Alarm times with $\beta =0.25$, for different $d$ and $\alpha$ -- Exponential case (Ex.1)}

{\scriptsize
\begin{tabular}{|c|ccccccccccc|}
\hline
 d &                                                                                                                  \multicolumn{ 11}{|c|}{$\alpha$} \\
  & {\bf 0.25} & {\bf 0.275} &  {\bf 0.3} & {\bf 0.325} & {\bf 0.35} & {\bf 0.375} &  {\bf 0.4} & {\bf 0.425} & {\bf 0.45} & {\bf 0.475} &  {\bf 0.5} \\

\hline

 0.60 &         NA &         NA &         NA &         NA &         NA &         NA &         NA &         NA &         NA &         NA &       NA \\
       0.65 &         NA &         NA &         NA &         NA &         NA &         NA &         NA &         NA &         NA &         NA &       0.44 \\

       0.7 &         NA &         NA &         NA &         NA &         NA &         NA &         NA &         NA &         NA &       0.42 &       0.29 \\

      0.75 &         NA &         NA &         NA &         NA &         NA &         NA &         NA &         NA &        0.4 &       0.27 &       0.19 \\

       0.8 &         NA &         NA &         NA &         NA &         NA &         NA &         NA &       0.39 &       0.27 &       0.18 &       0.12 \\

      0.85 &         NA &         NA &         NA &         NA &         NA &         NA &       0.39 &       0.27 &       0.18 &       0.11 &       0.06 \\

       0.9 &         NA &         NA &         NA &         NA &         NA &       0.42 &       0.27 &       0.18 &       0.11 &       0.05 &          0 \\

      0.95 &         NA &         NA &         NA &         NA &       0.45 &       0.27
&       0.18 &        0.1 &       0.05 &          0 &          0 \\

         1 &         NA &         NA &         NA &       0.49 &       0.29 &       0.18 &       0.11 &       0.05 &          0 &          0 &          0 \\

      1.05 &         NA &         NA &         NA &       0.33 &        0.2 &       0.11 &       0.05 &          0 &          0 &          0 &          0 \\

       1.1 &         NA &         NA &       0.37 &       0.22 &       0.12 &       0.06 &          0 &          0 &          0 &          0 &          0 \\

      1.15 &         NA &       0.46 &       0.26 &       0.14 &       0.06 &          0 &          0 &          0 &          0 &          0 &          0 \\
\hline
\end{tabular}
}
\end{center}

\begin{center}
{\bf \scriptsize Table 1.2:  Alarm times with $\beta =0.25$, for different $d$ and $\alpha$ -- Pareto case (Ex.2)}\\
{\scriptsize
\begin{tabular}{|c|cccccc|} \hline
         d  &                                                                                                     \multicolumn{ 6}{|c|}{$\alpha$} \\

          &  {\bf 0.35} & {\bf 0.375} &  {\bf 0.4} & {\bf 0.425} & {\bf 0.45} & {\bf 0.475} \\
\hline
      0.8 &         NA &         NA &         NA &         NA &         NA &         NA \\

      0.85 &                 NA &         NA &         NA &         NA &         NA &         NA \\

       0.9 &                NA &         NA &         NA &         NA &         NA &       0.39 \\

      0.95 &                  NA &         NA &         NA &         NA &       0.43 &       0.21 \\

         1 &                NA &         NA &         NA &         NA &       0.24 &       0.07 \\

      1.05 &               NA &         NA &         NA &       0.28 &       0.09 &          0 \\

       1.1 &                 NA &         NA &       0.35 &       0.13 &          0 &          0 \\

      1.15 &                 NA &       0.45 &       0.18 &       0.01 &          0 &          0 \\
\hline
\end{tabular}  }
\end{center}

\begin{center}
{\bf \scriptsize Table 1.3:  Alarm times with $\beta =0.05$, for different $d$ and $\alpha$ -- Logarithm case (Ex.3)}
{\scriptsize
\begin{tabular}{|c|cccccccccc|} \hline
       d    &                                                                   \multicolumn{ 10}{|c|}{$\alpha$} \\

        & {\bf 0.25} & {\bf 0.275} &  {\bf 0.3} & {\bf 0.325} & {\bf 0.35} & {\bf 0.375} &  {\bf 0.4} & {\bf 0.425} & {\bf 0.45} & {\bf 0.475} \\

      0.27 &         NA &         NA &         NA &         NA &         NA &         NA &         NA &         NA &         NA &         NA \\

      0.29 &         NA &         NA &         NA &         NA &         NA &         NA &         NA &         NA &         NA &       0.07 \\

      0.31 &         NA &         NA &         NA &         NA &         NA &         NA &         NA &       0.08 &       0.06 &       0.05 \\

      0.33 &         NA &         NA &         NA &         NA &         NA &         NA &       0.07 &       0.06 &       0.04 &       0.03 \\

      0.35 &         NA &         NA &         NA &         NA &       0.08 &       0.07 &       0.05 &       0.03 &       0.02 &          0 \\

      0.37 &         NA &         NA &         NA &       0.08 &       0.06 &       0.04 &       0.03 &       0.01 &          0 &          0 \\

      0.39 &         NA &         NA &       0.07 &       0.05 &       0.04 &       0.02 &       0.01 &          0 &          0 &          0 \\

      0.41 &         NA &       0.07 &       0.05 &       0.03 &       0.02 &          0 &          0 &          0 &          0 &          0 \\

      0.43 &       0.07 &       0.05 &       0.03 &       0.01 &          0 &          0 &          0 &          0 &          0 &          0 \\

      0.45 &       0.04 &       0.02 &       0.01 &          0 &          0 &          0 &          0 &          0 &          0 &          0 \\

      0.47 &       0.03 &       0.01 &          0 &          0 &          0 &          0 &          0 &          0 &          0 &          0 \\

      0.49 &          0 &          0 &          0 &          0 &          0 &          0 &          0 &          0 &          0 &          0 \\

\hline
\end{tabular}  }
\end{center}

\smallgap
{\bf Observations.} The results obtained from these simulations reconfirm the intuitive dependency of alarm time on $\alpha$, $\beta$ and $d$, viz.\:
\begin{itemize}
\item[-] for fixed $\alpha$ and $\beta$, the alarm time decreases with increase in $d$; for too small a choice of $d$, alarm never happens, and for too large a $d$, the alarm sounds instantaneously (alarm time =0). This is intuitively justified as the ruin probability increases with the future time horizon ($d$);
\item[-] for fixed $d$ and  $\beta$, the alarm time decreases with increase in $\alpha$. This is also intuitively clear as an increase in $\alpha$ amounts to be less restrictive i.e.\  more proactive in taking precautionary measures;
\item[-]  the impact of $\beta$ on the alarm time is interesting and perhaps less intuitive.  It appears that, if the alarm happens, then actually the timing of the alarm does not depend on $\beta$. Of course it is possible that for some small $\beta$, alarm does not sound ever while it does for large $\beta$; e.g.\
compare the alarm times for $d=1.15$, $\alpha=0.275$ from Table 1.1 and Table 1.2.
\end{itemize}

\subsubsection{How alarm time depends on Initial Capital}

\noindent It is intuitively obvious that the alarm time would increase with an increase in the initial capital. To observe the  pattern in a specific instance, we revert e.g. to the setting in Example 2, except the initial capital is varied from 0 to 120 and we observe the alarm times with $\beta = 0.225$, $ \alpha = 0.45$ and $ d= 1.0$. These alarm times are exhibited in  Figure 2.

\begin{figure}[H] 
  \begin{center}
  {\bf \scriptsize Figure 2:  First alarm time for various initial capital amounts -- Pareto case (Ex.2)}
\includegraphics[height=200pt,width=280pt]{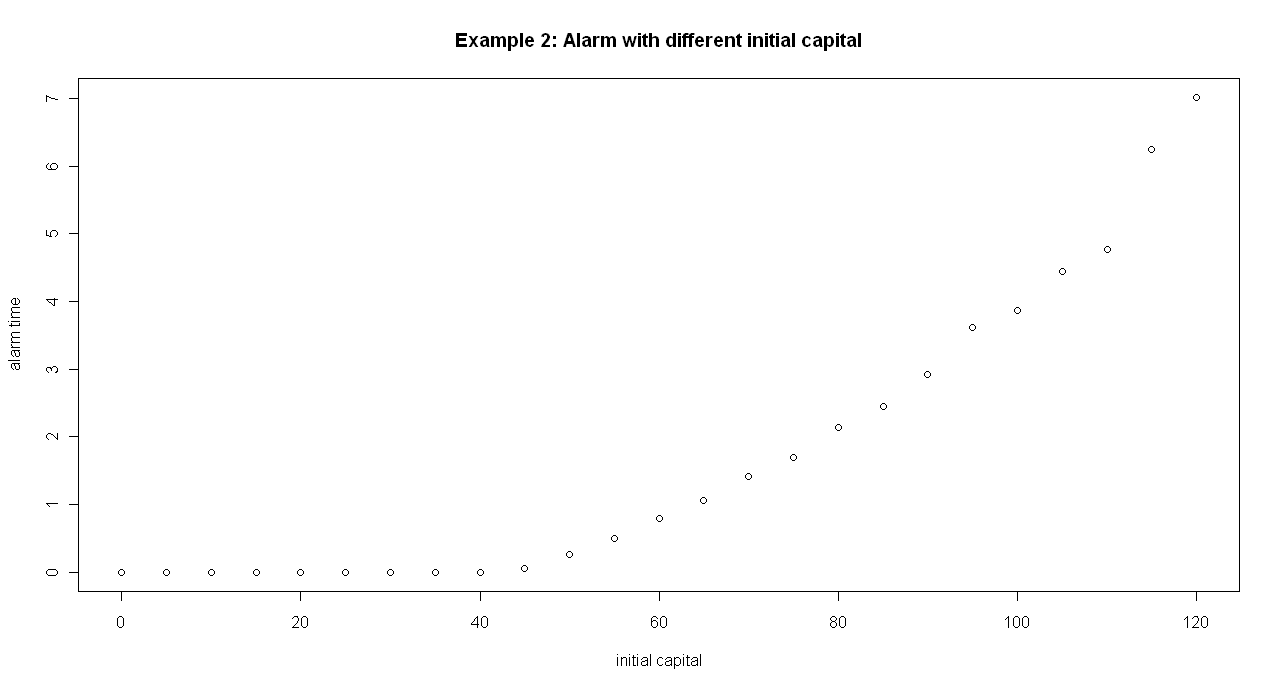}
 \end{center}
\end{figure}
\vspace{-0.5cm}

\noindent
As we can see the alarm sounds instantaneously unless the initial capital is at least 45 (possibly marginally less). For higher initial capital, alarm time is delayed almost linearly till the initial capital is 110. For even higher initial capital, the alarm time grows at a much higher rate and eventually alarm will not sound; indeed, there would be no ruin.\\
It is interesting to see how the infinite time (as well as possibly finite time, at different time points) ruin probabilities change with various amounts put as initial capital. Towards this, we compare the entire ruin-time distributions in Figure 3 for the various choices of the initial capital. We observe that the ruin-time increases with increase in initial capital which is reflected in the distribution getting more spread out with less peakedness.

\begin{figure}[H] 
\begin{center}
{\bf \scriptsize Figure 3: Density of Ruin-time with various Initial Capitals -- Pareto case (Ex.2)}
\includegraphics[height=200pt,width=400pt]{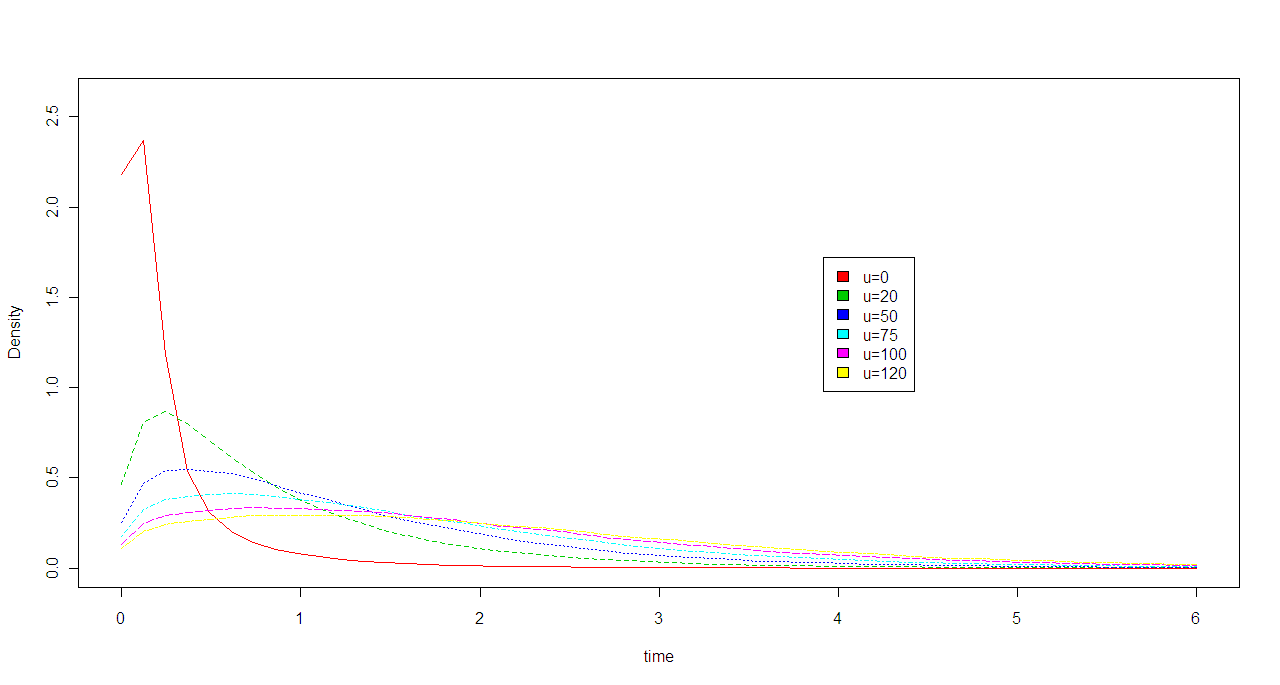}
\end{center}
\end{figure}
\noindent It may be more interesting to compare the probabilities of ruin before a given time directly. As expected the finite horizon ruin probabilities decrease with in initial capital, as it can be seen in Figure 4.
\begin{figure}[H] 
  \begin{center}
 {\bf \scriptsize Figure 4: Finite Horizon Ruin Probability for various initial capital -- Pareto case (Ex.2)}
\includegraphics[height=200pt,width=400pt]{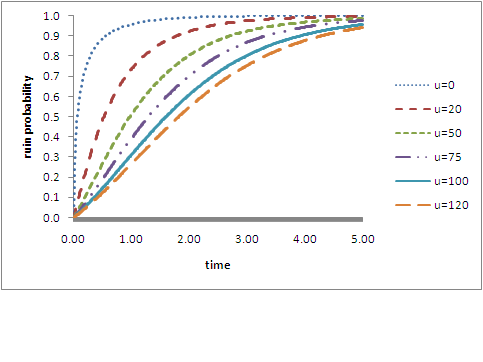}
 \end{center}
\end{figure}

\subsection{Multiple alarms leading to an alarm system with capital added at each alarm time}\label{multiple:alarm}

Suppose we have at time 0 a capital $u_0$.
To prevent the system from ruin, we propose to add a capital $u_1$ at time $A$. We generalize this procedure and define a system of alarm times  ${A_i}$, $i\ge 1$, adding a capital $u_i$ at every alarm time ${A_i}$. Note that the risk process $V(\cdot)$ gets then modified at each ${A_i}$, $i\ge 1$.

\smallgap Let us define the event $B_i$ for $i\ge 1$ as
 \begin{equation}\label{defBi}
B_{i}~:=~\bigcap_{n=1}^{i}\Big(T(A_{n-1},\sum_{m=0}^{n-1} u_m\big)>A_{n}\Big)
\end{equation}
(i.e. no ruin occurs up to alarm time $A_i$) and for the sake of consistency, set $A_0=0$ and $P(B_0)=1$.\\
Define for $i\ge 1$
\begin{eqnarray}\label{def_Ai}
{A_{i+1}}&=&\inf \Big\{ s > {A_i}: ~P\Big[ T({A_i},\sum_{n=0}^i u_n) \le s+d ~|~ \Big(T({A_i},\sum_{n=0}^i u_n) > s\Big) ~\cap~B_i~\Big] \geq 1 - \alpha \nonumber\\
&& \qquad\qquad\qquad\mbox{and}\quad  P\Big[ T({A_i},\sum_{n=0}^i u_n) > s~|~B_i~\Big] \geq 1 -\beta \Big\} \nonumber\\
&=&\!\!\inf \Big\{ s > {A_i}: ~\psi_{A_i}\Big(\sum_{n=0}^i u_n~,s+d~|~B_{i}\Big)-\psi_{A_i}\Big(\sum_{n=0}^i u_n~,s~|~B_{i}\Big)~\ge \nonumber\\
&& ~(1-\alpha)~
\bar\psi_{A_i}\Big(\sum_{n=0}^i u_n~,s~|~B_{i}\Big)
\qquad\mbox{and}\qquad \bar\psi_{A_i}\Big(\sum_{n=0}^i u_n~,s~|~B_{i}\Big) \geq 1 -\beta \Big\}.
\end{eqnarray}

\noindent This definition implies in particular that
\begin{equation}\label{ineqAi}
\psi_{A_i}\Big(\sum_{n=0}^i u_n~,s+d~|~B_{i}\Big)-\psi_{A_i}\Big(\sum_{n=0}^i u_n~,s~|~B_{i}\Big)~\ge~(1-\alpha)(1-\beta).
\end{equation}

\smallgap
\begin{remark}
For the sake of simplicity,  we keep here the same values for $\alpha$, $\beta$ {\color{blue}and $d$} in the  definition of successive alarm times; however, to generalize, one may consider a sequence of values $(\alpha_i)_{i\ge 1}$, $(\beta_i)_{i\ge 1}$ and $(d_i)_{i\ge 1}$ instead.\\
The capital $u_i$ added at each alarm time $A_i$ is given on purpose ; it corresponds in this way to the (available) amount that the company chooses to put at this time. The advantage is that it gives more flexibility to the company, but it also implies that the smallest the added capital are, the closest the alarm times may become. The same procedure is then repeated from any alarm time $A_i$ with the associated capital $\sum_{n=0}^i u_n$. \\
An alternative way would be to optimize the amount $u_i$ to be put at an alarm time $A_i$, such that the chance of ruin $P[T(A_i,\sum_{n=0}^i u_n)\le A_i+d]$ is very small. In this case, the procedure would start again at $A_i+d$ instead of at $A_i$. This optimized amount could also simply be used as an information for the company to be able to compare it with the amount they are ready to put in at this moment.
\end{remark}

\smallgap Let us now report the computational results in terms of timings of multiple alarms, in the context of Examples 1 to 3.

\smallgap {\it Revisit Example 1}. We consider the same risk process as in Example 1, except that the initial capital $u$ is changed to 17.5. As for the parameters in the definition of alarm, we take   $\beta = 0.225$, $ \alpha = 0.45$ and $ d= 1.0$. A fixed percentage of the initial capital is added to the capital at the sound of each alarm. In the following figure, we show the pattern of the first 30 alarm times, for the different choice of this percentage of the initial capital, that would be added to the system. As can be seen from the figure, when higher percentages of initial capital are added, alarms stop to sound after first few alarms.

\begin{figure}[H]
\begin{center}
 {\bf \scriptsize Figure 5:  The first 30 alarm times with different amounts added at each alarm -- Exponential case (Ex. 1)}
\includegraphics[height=200pt,width=280pt]{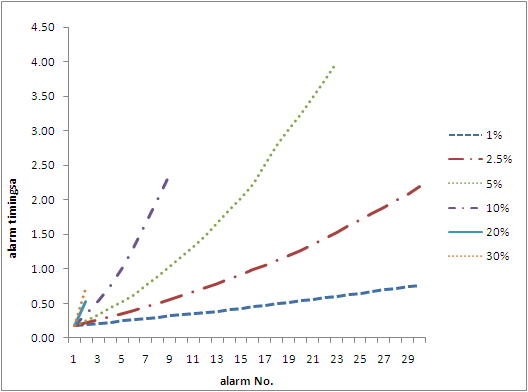}
\end{center}
\end{figure}

\smallgap {\it Revisit Example 2}. We now revisit the same risk process as in Example 2 and consider the alarm definition with $\beta = 0.225$, $ \alpha = 0.45$ and $ d= 1.0$. With 10\% of the initial capital $u=50$, being added to the system at each alarm, we get exactly 4 alarms occurring respectively
at $t=$ 0.29,	0.58,	0.91,	and 1.28.

\smallgap {\it Revisit Example 3}. With the setup of Example 3,
we choose  $\beta = 0.05$, $ \alpha = 0.4$ and $ d= 0.35$ as the parameters in the definition of alarm. Table 2A is with an initial capital of
 $u=50$,
while Table 2B is with $u=55$.

 \begin{table}[H]
\begin{center}
{\bf \scriptsize Table 2A: First 3 alarm times with various \% of $u_0=50$  added at each alarm -- Logarithm case (Ex.3)} \\[.1in]
{\scriptsize
\begin{tabular}{||c||c|c|c||}
\hline
\multicolumn{ 1}{|c|}{\% of u added} &                             \multicolumn{ 3}{|c|}{alarm times} \\
\multicolumn{ 1}{|c|}{at each alarm} &    alarm 1 &    alarm 2 &    alarm 3 \\
\hline
       2\% &       0.05 &       0.06 &       0.07 \\

       4\% &       0.05 &       0.06 &       0.08 \\

       6\% &       0.05 &       0.07 &       0.09 \\

       8\% &       0.05 &       0.08 &       0.11 \\

      10\% &       0.05 &       0.09 &       0.13 \\
\hline
\end{tabular}}
\end{center}
\end{table}
\vspace{-0.35in}

 \begin{table}[H]
\begin{center}
{\bf \scriptsize Table 2B: First 3 alarm times with various \% of $u_0=55$  added at each alarm -- Logarithm case (Ex.3)} \\[.1in]
{\scriptsize
\begin{tabular}{||c||c|c|c|c|c||}
\hline
\multicolumn{ 1}{|c|}{\% of u added} &                             \multicolumn{ 5}{|c|}{alarm times} \\
\multicolumn{ 1}{|c|}{at each alarm} &    alarm 1 &    alarm 2 &    alarm 3 &    alarm 4 &    alarm 5 \\
\hline
2.5\% &       0.08 &       0.09 &        0.1 &       0.12 &       0.13 \\

5\% &       0.08 &       0.11 &       0.13 &       0.15 &       0.18 \\

7.5\% &       0.08 &       0.12 &       0.15 &       0.19 &       0.22 \\

10\% &       0.08 &       0.13 &       0.17 &       0.22 &       0.27 \\

12.5\% &       0.08 &       0.14 &       0.19 &         NA &         NA \\

15\% &       0.08 &       0.15 &         NA &         NA &         NA \\

17.5\% &       0.08 &       0.16 &         NA &         NA &         NA \\

20\% &       0.08 &         NA &         NA &         NA &         NA \\
\hline
\end{tabular}}
\end{center}
\end{table}

\section{Effectiveness of Alarm System} \label{alarm:compare}

We consider the alarm system as an alternate strategy for having to put up an excessive initial capital to avoid ruin. This alternative strategy calls for starting with a comparatively modest initial capital, while additional amounts are to be added to the capital whenever the (suitably designed) alarms  go off. 
To draw a fair comparison, we consider the total capital to be equivalent under the two alternatives. Mathematically, as well as intuitively, it is clear that probability of ruin would be lower, at least in the short time horizon, when one puts the total amount at the very beginning; however we look for devising a suitable alarm system where the difference in ruin probability is nominal, especially when considering the time value of money (Net Present Value). Indeed, in such a case and under high enough interest regime, the ruin probability under alarm system may even become lower than the one without alarms. Of course, the time horizon over which the two systems are to be compared is the other yardstick that company would need to decide.  
More precisely, as described in the alarm system, whenever `alarm' sounds, a designed amount of capital may be added to the accumulation function to prevent from ruin, providing also a new risk process for which we would define the new alarm time.
Thus, in essence, we are proposing a piecewise linear accumulation function 
with discontinuities at the time point of alarms by upward parallel shift of the function.

\smallgap We want to compare two models, one named $M^k$, with $k\ge 1$ which may go to infinity, when considering an alarm system with $k$ alarm times as defined in (\ref{def_Ai}) and adding a certain amount $u_i$ at each $A_i$, and the other named $M_r$ when we do not take into account any alarm time but put an initial discounted amount $u_0+\sum_{i=1}^ke^{-rA_i}u_i$, where $r$ denotes a continuously compound interest rate.
We will compare the probability of ruin (in finite or infinite time) of those two models; to this aim, we propose a recursive method.\\
Note that if the NPC is violated, i.e. if $\theta\ge 0$, only the comparison of ruin probabilities in finite time will be of some interest.\\
In addition, we also explore the possibility of having infinite sequence of alarms ($k\to\infty$) and  compare it naturally with no-alarm system with infinite capital over infinite time horizon.

\subsection{Comparison of systems with or without an alarm: numerical illustration} \label{compare:numerical}

In this section we attempt to empirically verify if and/or when it is advantageous to have an alarm system with additional amounts being added at the sound of alarm as opposed to starting with higher initial capital (and no subsequent addition to the capital). To draw a fair comparison, it is imperative not only to subject both the risk processes to identical claim process but also ensure that the additional amounts provided for the non-alarm system is in congruence with the amounts added subsequently to the alarm-system. Since in our main approach, alarm times are non-random (parameters or fixed values), it is straightforward to consider the discounted amount as per the rate of interest $r$. \\
While  comparing performances  in the simulation set up,  we take several wide-ranging values of $r$,
Table 3 representing part of the result. It is important to note that while some of the choices for $r$ may appear to be unrealistic at first glance, it is not so because the unit of the time frame is unspecified in our framework. Consequently, these apparent high values of $r$ could be in fact quite reasonable if for example unit time ( $t=1$) corresponds to a long period like 10 years.

The comparison is carried out in the framework of Example 2, where the alarm system starts with an initial capital of $u=50$ and additional 10\% (=5) of the initial capital being added to the system at each of the alarm times. In the framework, we allowed for as many alarms as required and alarm goes off only on four occasions, viz.\ at $t$= 0.29,	0.58,	0.91	and 1.28 with
$\alpha= 0.45$, $\beta=	0.225$ and $d=	1$.
Some of the key survival probabilities $P[T \geq t]$ are reported in Table 3. For more complete comparison,
the probabilities of survivals up to  different time points [viz.\  survival function of the corresponding ruin times] of the systems are shown in Figure 6.

\begin{table}[H]
\begin{center} {\bf \scriptsize Table 3: Comparing Survival Probabilities between Systems without \& with alarm -- Pareto case (Ex.2)}
{\scriptsize
\begin{tabular}{|c|c|ccccccccc|} \hline

           & \multicolumn{ 1}{|c|}{} &                       \multicolumn{ 8}{|c}{NO alarm system with equivalent initial capital $u_0 (M_r)= u_0 + \sum_i e^{-r A_i} u_i $} &            \\
 & \multicolumn{ 1}{|c|}{$\qquad\qquad r$} &     0\% &       10\% &       30\% &       50\% &      100\% &      150\% &      200\% &      500\% &   $\infty$  \\ \hline

           &     $\qquad\quad u_0(M_r)$ &     70.000 &     68.540 &     65.996 &     63.875 &     59.944 &     57.341 &     55.564 &     51.509 &     50.000   \\
           \hline

          $t$ & $P(T^{M^A} \ge t)\qquad$ &                                                                      \multicolumn{ 9}{|c|}{$P[T^{M_r} \geq t]$}    \\ \hline

      0.01 & {\bf 0.995} &      0.996 &      0.996 &      0.996 &      0.996 &      0.996 &      0.996 &      0.996 &      0.995 &      0.995             \\

       0.1 & {\bf 0.950} &      0.964 &      0.963 &      0.962 &      0.961 &      0.958 &      0.956 &      0.955 &      0.952 &      0.950             \\

       0.2 & {\bf 0.896} &      0.924 &      0.922 &      0.919 &      0.917 &      0.911 &      0.908 &      0.905 &      0.899 &      0.896             \\
$A_1$= 0.29 & {\it {\bf 0.847}} & {\it 0.887} & {\it 0.885} & {\it 0.881} & {\it 0.877} & {\it 0.870} & {\it 0.864} & {\it 0.861} & {\it 0.851} & {\it 0.847} \\

       0.3 & {\bf 0.842} &      0.883 &      0.880 &      0.876 &      0.872 &      0.865 &      0.859 &      0.855 &      0.845 & {\bf 0.841}             \\

       0.4 & {\bf 0.794} &      0.840 &      0.837 &      0.832 &      0.826 &      0.816 &      0.809 &      0.804 &      0.791 & {\bf 0.785}             \\

       0.5 & {\bf 0.743} &      0.796 &      0.792 &      0.785 &      0.779 &      0.767 &      0.758 &      0.751 & {\bf 0.736} & {\bf 0.729}             \\

$A_2$= 0.58 & {\it {\bf 0.703}} & {\it 0.760} & {\it 0.756} & {\it 0.749} & {\it 0.742} & {\it 0.728} & {\it 0.718} & {\it 0.711} & {\it {\bf 0.693}} & {\it {\bf 0.686}} \\

      0.75 & {\bf 0.632} &      0.687 &      0.682 &      0.673 &      0.664 &      0.648 &      0.636 & {\bf 0.627} & {\bf 0.607} & {\bf 0.599}             \\
$A_3$= 0.91  &{\it {\bf 0.566}} & {\it 0.621} & {\it 0.615} & {\it 0.605} & {\it 0.596} & {\it 0.578} & {\it {\bf 0.565}} & {\it {\bf 0.555}} & {\it {\bf 0.533}} & {\it {\bf 0.524}} \\

         1 & {\bf 0.536} &      0.584 &      0.578 &      0.568 &      0.558 &      0.540 & {\bf 0.527} & {\bf 0.517} & {\bf 0.494} & {\bf 0.486}             \\

      1.25 & {\bf 0.451} &      0.490 &      0.484 &      0.472 &      0.462 & {\bf 0.443} & {\bf 0.430} & {\bf 0.420} & {\bf 0.397} & {\bf 0.388}             \\
$A_4$= 1.28 & {\it {\bf 0.442}} & {\it 0.481} & {\it 0.474} & {\it 0.463} & {\it 0.453} & {\it {\bf 0.434}} & {\it {\bf 0.421}} & {\it {\bf 0.410}} & {\it {\bf 0.387}} & {\it {\bf 0.379}} \\

       1.5 & {\bf 0.384} &      0.409 &      0.403 &      0.391 & {\bf 0.380} & {\bf 0.361} & {\bf 0.349} & {\bf 0.339} & {\bf 0.318} & {\bf 0.310}             \\

      1.75 & {\bf 0.321} &      0.337 &      0.331 & {\bf 0.320} & {\bf 0.311} & {\bf 0.293} & {\bf 0.281} & {\bf 0.272} & {\bf 0.253} & {\bf 0.246}             \\

         2 & {\bf 0.265} &      0.276 &      0.270 & {\bf 0.260} & {\bf 0.252} & {\bf 0.235} & {\bf 0.225} & {\bf 0.217} & {\bf 0.200} & {\bf 0.194}             \\

       2.5 & {\bf 0.177} &      0.183 &      0.178 & {\bf 0.171} & {\bf 0.164} & {\bf 0.151} & {\bf 0.143} & {\bf 0.137} & {\bf 0.125} & {\bf 0.120}            \\

         3 & {\bf 0.116} &      0.119 & {\bf 0.116} & {\bf 0.110} & {\bf 0.105} & {\bf 0.096} & {\bf 0.090} & {\bf 0.086} & {\bf 0.078} & {\bf 0.074}             \\

       3.5 & {\bf 0.076} &      0.077 & {\bf 0.075} & {\bf 0.070} & {\bf 0.067} & {\bf 0.060} & {\bf 0.056} & {\bf 0.053} & {\bf 0.048} & {\bf 0.045}             \\

         4 & {\bf 0.049} &      0.050 & {\bf 0.048} & {\bf 0.045} & {\bf 0.042} & {\bf 0.038} & {\bf 0.035} & {\bf 0.033} & {\bf 0.029} & {\bf 0.028}             \\

         5 & {\bf 0.021} &      0.021 & {\bf 0.020} & {\bf 0.019} & {\bf 0.018} & {\bf 0.016} & {\bf 0.015} & {\bf 0.014} & {\bf 0.012} & {\bf 0.011}             \\

         6 & {\bf 0.009} &      0.009 & {\bf 0.008} & {\bf 0.008} & {\bf 0.007} & {\bf 0.006} & {\bf 0.006} & {\bf 0.005} & {\bf 0.005} & {\bf 0.005}             \\

         7 & {\bf 0.004} &      0.004 & {\bf 0.004} & {\bf 0.003} & {\bf 0.003} & {\bf 0.003} & {\bf 0.003} & {\bf 0.002} & {\bf 0.002} & {\bf 0.002}             \\
 \hline
\end{tabular}    }
\end{center}
\end{table}

\begin{figure}[H] 
\begin{center}
  {\bf \scriptsize Figure 6: Survival function with or without alarm (different ROI)-- Pareto case (Ex.2)}
\includegraphics[height=200pt,width=350pt]{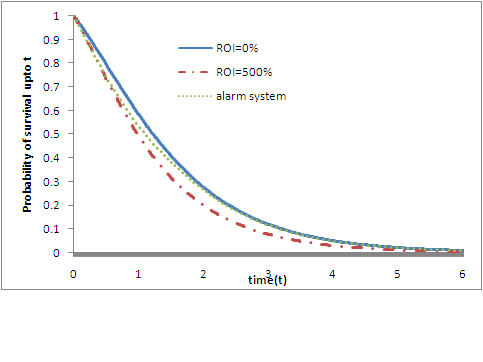}
 \end{center}
\end{figure}
\vspace{-0.5cm}

\noindent These table and figure clearly bring out the advantages of the alarm system. For example, if one considers the survival at time = 3 to be the reference frame, then it is beneficial to have the alarm system as opposed to having additional capital in the beginning as long as the rate of interest is 10\% or higher. Of course, if one considers a very short time window, the conclusion would be otherwise.

\smallgap
We refer to \cite{wpDK} for another numerical illustration between alarm and no-alarm systems, this time  in the setup of Example 1 (Exponential claims), where there is a total of 15 alarms.\\
In search for comparison between no-alarm and alarm systems in completely general situation, we target to obtain analytical upper and lower bounds of the difference in survival probabilities under the two systems. This is made possible thanks to a recursive approach taken up in the next subsection, a direct method leading to rougher bounds.

\subsection{Recursive method} \label{compare:analytical}

This method consists in introducing $k-1$ models having specific number $1$ to $k-1$ alarms in order to come back to a comparison between a one alarm model and a model without alarm but adjusted initial amount.\\
To enable a fair comparison across the models, we consider the amount being put at the time of the alarm to be such that amounts are equivalent across the model considering the time value of money. To be specific, the model $M^k$ (the final model) calls for starting with initial capital $u_0$ and putting an amount $u_i$ at the time of the $i$-th alarm, i.e.\ ${A_i}$, $\forall i=1,\ldots,k$.
On the other hand, the model preceding it, $M^{k-1}$ allows for capital change only at the time of the first $k-1$ alarms and the amount being put at all but the last of them coincides with the same for model $M^k$. Hence, at the time of the $(k-1)-$th alarm in model $M^{k-1}$, the amount put is given by
$$ u_{k-1} + u_k \times e^{-r({A_k}-{A_{k-1}})}.$$ 
Subsequent models are defined similarly. To consider a complete mathematical framework, let us identify ${A_0}=0$ (0-th alarm time) as the starting time of the process. Note that the stochastic process behind all the models remains the same as $R_t$; however the ruin or otherwise as per model $M^i$ at time $t$, depends on the level applicable  at that time, dictated by the amounts added to the system, as per the model
specifications. For $i=0,1,\ldots,k$, referring to Model $M^i$, these levels at time $t$, denoted by $l^i_t$, are given by
 \begin{equation}
 l^i_t = \left\{   \begin{array}{lc}
\displaystyle{\sum_{j=0}^m} u_j & \mbox{\rm if } t \in (A_m, A_{m+1}] \;  \mbox{\rm for some } m < i, \\
 \displaystyle{\sum_{j=0}^{i}} u_j  + \displaystyle{\sum_{j=i+1}^k} u_j \times  e^{-r({A_j}-{A_i})}=l^i_{A_{i+1}}  & \mbox{\rm for } t > A_i  \end{array}
 \right. \label{level_model}
 \end{equation}
Thus, for $i=0,1,\ldots,k-1$, the levels from the  successive models may be compared as:
\begin{equation}
 l^i_t = l^{i+1}_t,~ \forall t \le {A_i}; \qquad  l^i_t > l^{i+1}_t,~ \forall t \in (A_i,  A_{i+1}];~ \qquad
 l^i_t < l^{i+1}_t,~ \forall t > {A_{i+1}}. \label{level:compare}
 \end{equation}
Note that we set $A_{k+1}= \infty$ and the model $M^0$ corresponds to $M_r$ with level
$$l_t^0=u_0+\sum_{j=1}^k u_j~e^{-r A_j}, \qquad \forall t>A_0=0. $$
The comparison across the consecutive levels may be better understood through the following figure:
\begin{figure}[H] 
  \begin{center}
    \includegraphics[height=250pt,width=250pt]{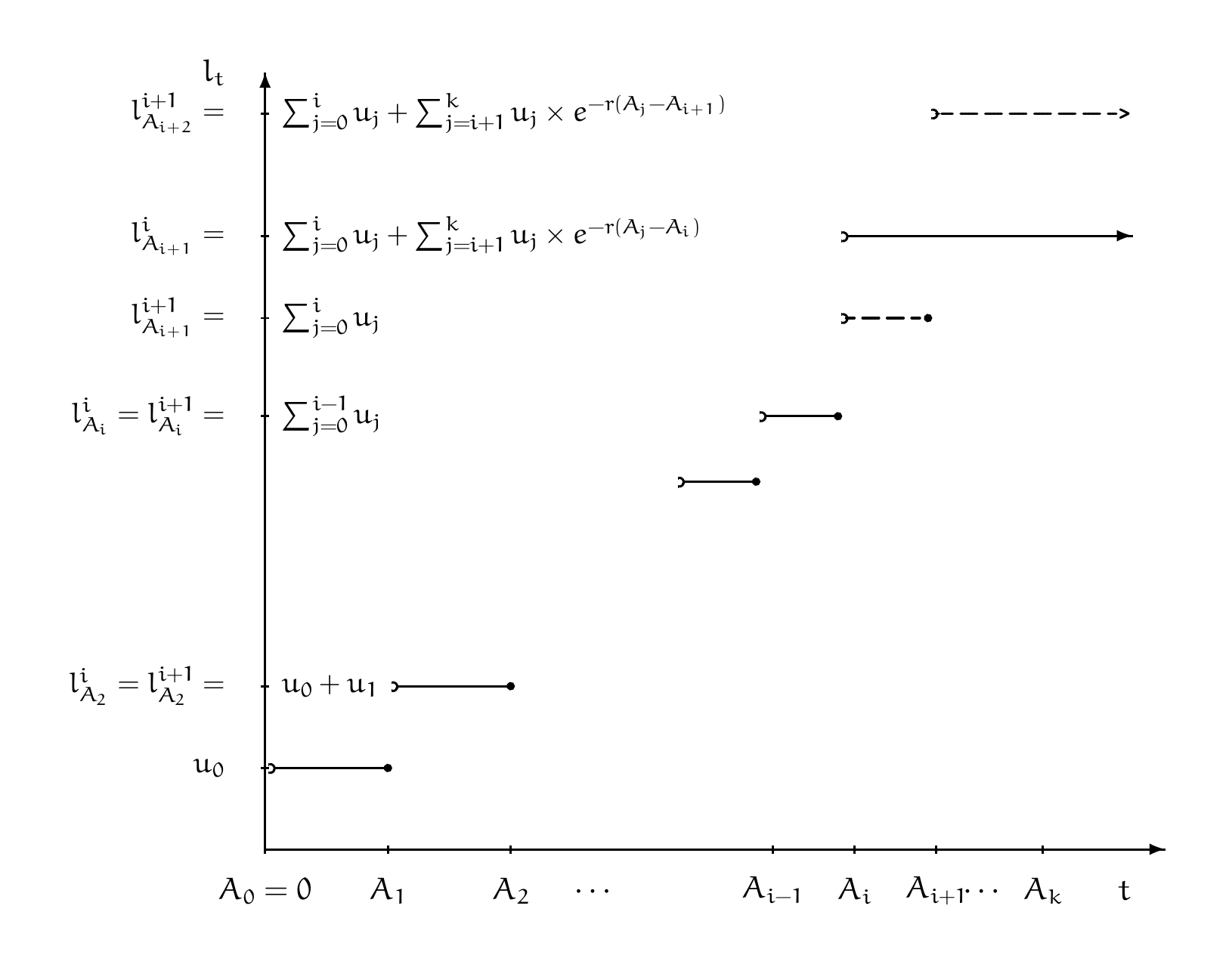}\\[-.25in]
    {\bf \scriptsize Figure 7: Capital levels under $M^i$ (solid line) and $M^{i+1}$ (dashed line, when the capital is different after $A_i$)}
\end{center}
\end{figure}

\subsubsection{Evaluation in finite time}

\noindent We want to evaluate
$$
\Delta(t)~=\Delta_{k,r}(t)~= ~P(T^{M^k} \le t)~-~ P(T^{M_r}\le t), \quad t>0.
$$
\underline{If $t\le A_1$}, then
$\displaystyle ~\big(T^{M^k} \le t\big)~=~\big(T(u_0)\le t\big)~$ and so
\begin{equation}\label{finiteBeforeAlarm}
\Delta(t)~=~\psi(u_0,t)-\psi(u_0+\sum_{j=1}^k e^{-rA_j}u_j,t).
\end{equation}
Since $\psi(u,t)$ is a non-decreasing function of the time $t$ and a non-increasing function of the capital $u$, using the definition (\ref{def_Ai}) of the alarm times $A_i$, we obtain
$$
0 <\Delta(t) \le \psi(u_0,A_1)-\psi(u_0+\sum_{j=1}^k e^{-rA_j}u_j,t)
\le \beta-\psi(u_0+e^{-rt}\sum_{j=1}^k u_j,t)\le \beta.
$$
In this case, model $M_r$ has edge over the model $M^k$, as is intuitively obvious, with the advantage reducing with increase in time. Nevertheless the upper bound is $\beta$, hence can be small enough.\\

\noindent \underline{For $t> A_1$}, we use the recursive approach to compute $\Delta(t)$ according to
$$
\Delta(t)~=~\sum_{j=0}^{k-1} \big(P[T^{M^{j}}>t]-P[T^{M^{j+1}}>t]\big).
$$
We know that there exists $i$ with $2\le i\le k+1$ such that $t\in (A_{i-1},A_{i}]$ (with $A_{k+1}=\infty$).\\
So from now on, we consider that $t\in (A_{i-1},A_{i}]$, for $2\le i\le k+1$.

\smallgap For $i\le k$, $\Delta(t)$ can be written as
\begin{eqnarray*}\label{delta_t_sum}
\Delta(t)&=&\sum_{j=0}^{i-1} \big(P[T^{M^{j}}>t]-P[T^{M^{j+1}}>t]\big)~+~
\sum_{j=i}^{k-1} \big(P[T^{M^{j}}>t]-P[T^{M^{j+1}}>t]\big)\nonumber\\
&=& \sum_{j=0}^{i-1} \big(P[T^{M^{j}}>t]-P[T^{M^{j+1}}>t]\big)
\end{eqnarray*}
since the models $M^j$ and $M^{j+1}$ have the same levels of reference until time $A_j$, hence until $t$, for any $j\ge i$; \\
for $t>A_k$, $\displaystyle \Delta(t)=\sum_{j=0}^{k-1} \big(P[T^{M^{j}}>t]-P[T^{M^{j+1}}>t]\big)$.\\
Therefore, for $2\le i\le k+1$, we can write
\begin{eqnarray}\label{delta_t_sum}
\Delta(t)&=& \sum_{j=0}^{i-2} \Big(P[T^{M^{j}}>t]-P[T^{M^{j+1}}>t]\Big)~+~1_{(i\le k)}\big(P[T^{M^{i-1}}>t]-P[T^{M^{i}}>t]\big).
\end{eqnarray}

\smallgap Let us start to evaluate the second term on the RHS of (\ref{delta_t_sum}).\\
Noticing that for any $j$, the two events (no ruin for $M^j$ up to $A_j$) and (no ruin for $M^{j+1}$ up to $A_j$) coincide whenever $t\le A_j$ because of (\ref{level:compare}), we can write for $i\le k$
\begin{eqnarray}\label{delta_t_iterm}
P[T^{M^{i-1}}>t]-P[T^{M^{i}}>t]
&=& P(B_{i-1})~\Big[\psi_{A_{i-1}}\big(l^{i}_{A_i},t~|~B_{i-1}\big) ~-~\psi_{A_{i-1}}\big(l^{i-1}_{A_i},t~|~B_{i-1}\big)\Big]
\end{eqnarray}
the event $B_j$ being defined in (\ref{defBi}),\\
or, in a similar way, using (\ref{psiconditio}),
\begin{eqnarray}\label{delta_t_itermBis}
P[T^{M^{i-1}}>t]-P[T^{M^{i}}>t]
&=& \int_{-\infty}^{l^{i-1}_{A_{i-1}}}
\Big\{ \psi\big(l^{i}_{A_i}-x~,~ t-A_{i-1} \big) ~-~\psi\big(l^{i-1}_{A_i}-x, t-A_{i-1}\big)\Big\}\nonumber\\
&&\qquad\qquad\qquad \qquad\qquad P\big[B_{i-1}~|~R_{A_{i-1}}=x \big]~dF_{R_{A_{i-1}}}(x).\quad
\end{eqnarray}
An upper bound can then be deduced as
\begin{eqnarray}
P[T^{M^{i-1}}>t]-P[T^{M^{i}}>t]&\le &
P\big(B_{i-1}\big)\max_{x\le l^{i-1}_{A_{i-1}}}
\!\!\Big( \psi\big(l^{i}_{A_i}-x~,~ t-A_{i-1} \big)~-~\psi\big(l^{i-1}_{A_i}-x, t-A_{i-1}\big)\Big)\qquad\label{upperBdTerm_i}\\
&\le&P\big(B_{i-1}\big)\psi\big(l^{i}_{A_i}-l^{i-1}_{A_{i-1}}~,~ t-A_{i-1} \big)=P\big(B_{i-1}\big)\psi\big(u_{i-1}~,~ t-A_{i-1} \big) \nonumber
\end{eqnarray}
whereas a lower bound is given by
\begin{eqnarray}\label{lowerBdTerm_i}
P[T^{M^{i-1}}>t]-P[T^{M^{i}}>t]&\ge &
P\big(B_{i-1}\big)\min_{x\le l^{i-1}_{A_{i-1}}}
\!\!\Big( \psi\big(l^{i}_{A_i}-x~,~ t-A_{i-1} \big)~-~\psi\big(l^{i-1}_{A_i}-x, t-A_{i-1}\big)\Big).\qquad
\end{eqnarray}

\smallgap Now let us consider the first term on the RHS of (\ref{delta_t_sum}).\\
For any $0\le j\le i-2$, since $t> A_{i-1}$, we have
\begin{eqnarray}\label{deltaj(x,t)}
&&P[T^{M^{j}}>t]-P[T^{M^{j+1}}>t] \nonumber\\
&=&\int_{-\infty}^{~l_{A_j}^j} \Big(~ P[(\mbox{no ruin for}~ M^{j}~\mbox{up to}~ A_j)\cap (\mbox{no ruin for}~ M^{j}~\mbox{between} ~A_j~\mbox{and} ~t)~|~R_{A_j}=x]~-\nonumber\\
&& ~P[(\mbox{no ruin for}~ M^{j+1}~\mbox{up to}~ A_j)\cap (\mbox{no ruin for}~ M^{j+1}~\mbox{between} ~A_j~\mbox{and} ~t)~|~R_{A_j}=x]~\Big)dF_{R_{A_j}}(x)  \nonumber\\
&=& \int_{-\infty}^{~l_{A_j}^j} P\big[B_j~|~R_{A_j}=x\big]~\Delta^j (x,t)~dF_{R_{A_j}}(x)
\end{eqnarray}
\begin{eqnarray*}
&\mbox{with}&\quad \Delta^j(x,t) :=
P\Big[\Big(T\big(A_j,l^j_{A_{j+1}}\big)> t\Big)~|~B_j~\cap~(R_{A_j}=x)\Big]~-\\
&& \qquad\qquad\qquad P\Big[\Big(T\big(A_j,l^{j+1}_{A_{j+1}}\big)> A_{j+1}\Big)\cap
\Big(T\big(A_{j+1},l^{j+1}_{A_{j+2}}\big)> t\Big)
~|~B_j~\cap~(R_{A_j}=x)\Big]
\end{eqnarray*}
\noindent Note that $\Delta^j(x,t)$, for $0\le j \le i-2 $, corresponds to the difference of ruin probabilities for finite time $t>A_{j+1}$ between two models, one without alarm and the other with one alarm time $A_{j+1}$, when taking the time origin at $A_j$ with the value $R_{A_j}=x$, and the associated amounts: $\quad l^{j}_{A_{j+1}}=\sum_{n=0}^{j}u_n+\sum_{n=j+1}^{k}u_n e^{-r(A_n-A_{j})}$ and
$l^{j+1}_{A_{j+1}}=\sum_{n=0}^{j}u_n$ as initial amounts for the model without alarm and with an alarm respectively, and $l^{j+1}_{A_{j+2}}=\sum_{n=0}^{j+1}u_n+\sum_{n=j+2}^{k}u_n e^{-r(A_n-A_{j+1})}$ as the amount at $A_{j+1}$ for the alarm model.\\
Using the independence and stationarity of the increments of $R$, and (\ref{psiconditio}), we can write
\begin{eqnarray}\label{exactDelta_iEquiv(x,t)}
\Delta^j(x,t)
&=& \bar\psi\big(l^j_{A_{j+1}}-x, t-A_j\big) ~-~
\int_{-\infty}^{~l_{A_{j+1}}^{j+1}} \bar\psi\big(l^{j+1}_{A_{j+2}}-z~,~t-A_{j+1}\big) \times \\
&&\qquad\quad \bar\psi\Big(l^{j+1}_{A_{j+1}}-x~,~A_{j+1}-A_j ~|~R_{A_{j+1}-A_j}=z-x\Big)~ ~dF_{R_{A_{j+1}-A_j}}(z)\nonumber
\end{eqnarray}
or, equivalently,
\begin{eqnarray}\label{exactDelta_iBis(x,t)}
&& \Delta^j (x,t) = \psi\big(l^{j+1}_{A_{j+2}}-x,t-A_j\big)-
\psi\big(l^{j}_{A_{j+1}}-x,t-A_j\big) + \\
&&\int_{-\infty}^{l^{j+1}_{A_{j+2}}-x} \bar\psi\big(l^{j+1}_{A_{j+2}}-x-y,t-A_{j+1}\big)\Big[\psi\big(l^{j+1}_{A_{j+1}}-x,A_{j+1}-A_{j}~|~R_{A_{j+1}-A_j}=y\big)\nonumber\\
&&\qquad\qquad\qquad\qquad\quad -~\psi\big(l^{j+1}_{A_{j+2}}-x,A_{j+1}-A_{j}~|~R_{A_{j+1}-A_j}=y\big)\Big]dF_{R_{A_{j+1}-A_j}}(y).\nonumber
\end{eqnarray}

\smallgap Therefore, combining (\ref{delta_t_sum}) with (\ref{delta_t_itermBis}), (\ref{deltaj(x,t)}), (\ref{exactDelta_iEquiv(x,t)}) (or (\ref{exactDelta_iBis(x,t)}), respectively), the comparison between the two models can be evaluated for $t\in (A_{i-1},A_i]$, $2\le i\le k+1$, according to
\begin{eqnarray}\label{delta_t_sumFinal}
\Delta(t)&=&
\sum_{j=0}^{i-2} \int_{-\infty}^{~l_{A_j}^j}P\big[B_j~|~R_{A_j}=x\big]~
\Big\{\bar\psi\big(l^j_{A_{j+1}}-x, t-A_j\big) ~-~
\int_{-\infty}^{~l_{A_{j+1}}^{j+1}} \bar\psi\big(l^{j+1}_{A_{j+2}}-y~,~t-A_{j+1}\big) \times\\
&& \qquad\qquad\qquad\bar\psi\Big(l^{j+1}_{A_{j+1}}-x~,~A_{j+1}-A_j ~|~R_{A_{j+1}-A_j}=y-x\Big)~ ~dF_{R_{A_{j+1}-A_j}}(y)\Big\}~dF_{R_{A_j}}(x)\nonumber\\
&+&\!\!\! 1_{(i\le k)}\int_{-\infty}^{l^{i-1}_{A_{i-1}}}\Big\{\psi\big(l^{i}_{A_i}-x~,~ t-A_{i-1}\big) ~-~ \psi\big(l^{i-1}_{A_i}-x~,~ t-A_{i-1}\big)\Big\}P\big[B_{i-1}~|~R_{A_{i-1}}=x\big]~dF_{R_{A_{i-1}}}(x)\nonumber
\end{eqnarray}
or equivalently
\begin{eqnarray}\label{delta_t_sumFinalBis}
\Delta(t)\!\!\!&=& -\sum_{j=0}^{i-2}\int_{-\infty}^{~l_{A_j}^j}
\left\{\psi\big(l^{j}_{A_{j+1}}-x~,~t-A_j\big)~-~\psi\big(l^{j+1}_{A_{j+2}}-x~,~t-A_j\big)\right\}~
P\big[B_j~|~R_{A_j}=x\big]~dF_{R_{A_j}}(x) \nonumber\\
\!\!\!&+& \sum_{j=0}^{i-2}\int_{-\infty}^{~l_{A_j}^j}P\big[B_j~|~R_{A_j}=x\big]\int_{-\infty}^{l^{j+1}_{A_{j+2}}-x} \Big[\psi\big(l^{j+1}_{A_{j+1}}-x,A_{j+1}-A_{j} ~|~R_{A_{j+1}-A_j}=y\big)~ - \\
\!\!\!&&\!\!\! \psi\big(l^{j+1}_{A_{j+2}}-x,A_{j+1}-A_{j}~|~R_{A_{j+1}-A_j}=y\big)\Big]\bar\psi\big(l^{j+1}_{A_{j+2}}-x-y~,~t-A_{j+1}\big)~dF_{R_{A_{j+1}-A_j}}(y)~~dF_{R_{A_j}}(x)\nonumber\\
&+&\!\!\! 1_{(i\le k)}\int_{-\infty}^{l^{i-1}_{A_{i-1}}}\Big\{\psi\big(l^{i}_{A_i}-x~,~ t-A_{i-1}\big) ~-~ \psi\big(l^{i-1}_{A_i}-x~,~ t-A_{i-1}\big)\Big\}~P\big[B_{i-1}~|~R_{A_{i-1}}=x\big]~dF_{R_{A_{i-1}}}(x).\nonumber
\end{eqnarray}

\smallgap Those results can be extended (considering $i=k+1$ in (\ref{delta_t_sumFinal}) or (\ref{delta_t_sumFinalBis})) when taking the limit as $t\to\infty$ to express the ultimate ruin probabilities as:
\begin{eqnarray}\label{generalDelta}
\Delta &=& \sum_{j=0}^{k-1} \int_{-\infty}^{~l_{A_j}^j} \Delta^j (x)~P\big[B_j~|~(R_{A_j}=x)\big]~dF_{R_{A_j}}(x) \qquad\mbox{with}
\end{eqnarray}
\begin{eqnarray}\label{exactDelta_j}
 \Delta^j(x) &=& \bar\psi\big(l^j_{A_{j+1}}-x\big) ~-~
\int_{-\infty}^{~l_{A_{j+1}}^{j+1}} \bar\psi\big(l^{j+1}_{A_{j+2}}-y\big) \times\nonumber\\
&&\qquad\quad \bar\psi\Big(l^{j+1}_{A_{j+1}}-x~,~A_{j+1}-A_j ~|~R_{A_{j+1}-A_j}=y-x\Big)~ ~dF_{R_{A_{j+1}-A_j}}(y)
\end{eqnarray}
or equivalently
\begin{eqnarray}\label{exactDelta_jBis}
\Delta^j (x) &=&
\int_{-\infty}^{l^{j+1}_{A_{j+2}}-x} \bar\psi\big(l^{j+1}_{A_{j+2}}-x-y\big)~\Big[\psi\big(l^{j+1}_{A_{j+1}}-x,A_{j+1}-A_{j} ~|~R_{A_{j+1}-A_j}=y\big)\nonumber\\
&&\qquad\qquad -~\psi\big(l^{j+1}_{A_{j+2}}-x,A_{j+1}-A_{j} ~|~R_{A_{j+1}-A_j}=y\big)\Big]~dF_{R_{A_{j+1}-A_j}}(y) \nonumber\\
&-& \Big(\psi\big(l^{j}_{A_{j+1}}-x\big) ~-~ \psi\big(l^{j+1}_{A{j+2}}-x\big)\Big).
\end{eqnarray}

\smallgap $\triangleright$ {\it Bounds for} $\Delta(t)$, $t\in (A_{i-1},A_{i}]$ ($2\le i\le k+1$).\\

\noindent Notice, before starting, that we have for any $t>A_j$ and any amount $u$,
\begin{eqnarray}\label{psiAi|Bi}
\psi_{A_j}(u,t~|~B_j)
&=&\int_{-\infty}^{u}\psi(u-x,t-A_j)~P\Big[B_j~|~R_{A_j}=x\Big]~dF_{R_{A_j}}(x),
\end{eqnarray}
from which we deduce another expression for (\ref{def_Ai}), namely
\begin{eqnarray}\label{def_AiBis}
{A_{j+1}}\!\!&=&\!\!\inf \Big\{ s > {A_j}:\int_{-\infty}^{l^j_{A_j}} \Big[\psi\Big(\sum_{n=0}^j u_n -x,s-A_j+d\Big)-\psi\Big(\sum_{n=0}^j u_n -x,s-A_j\Big)\Big]~P\Big[B_j~|~R_{A_j}=x\Big]~dF_{R_{A_j}}(x) \nonumber\\
&&\qquad\qquad \qquad \qquad \ge~(1-\alpha)~\int_{-\infty}^{l^j_{A_j}}
\bar\psi\Big(\sum_{n=0}^j u_n~-x~,~s-A_j\Big)~P\Big[B_j~|~R_{A_j}=x\Big]~dF_{R_{A_j}}(x)\nonumber\\
&& \qquad\qquad\mbox{and}\quad  \int_{-\infty}^{l^j_{A_j}}
\bar\psi\Big(\sum_{n=0}^j u_n~-x,s-A_j\Big)~P\Big[B_j~|~R_{A_j}=x\Big]~dF_{R_{A_j}}(x)~\ge (1 -\beta)~P(B_j)~\Big\}.
\end{eqnarray}
Therefore we will also often use the following approximation:
\begin{equation}\label{def_AiBisApprox}
\int_{-\infty}^{l^j_{A_j}}
\bar\psi\Big(l^{j+1}_{A_{j+1}}~-x,A_{j+1}-A_j\Big)~P\Big[B_j~|~R_{A_j}=x\Big]~dF_{R_{A_j}}(x)~\simeq (1 -\beta)~P(B_j).
\end{equation}

\noindent After this preamble, first let us compute $P(B_j)$, for $1\le j\le k$, since it will be needed to evaluate the bounds of $\Delta(t)$. \\
Definitions (\ref{defBi}) and (\ref{def_Ai}) induce the recursive relation
$$
P(B_{j+1})~=~ P\big[T\big(A_{j}, \sum_{n=0}^{j} u_n\big)> A_{j+1}~|~B_j\big]~P(B_j)~\ge ~(1-\beta)P(B_{j})
$$
which implies, for $j\ge 1$,
\begin{equation}\label{PBi}
P(B_j)~\ge ~ (1-\beta)^{j-1} P(B_1)~\ge ~(1-\beta)^j;
\end{equation}
nevertheless, because of (\ref{def_Ai}), we will often consider that
\begin{equation}\label{PBiApprox}
P(B_j)~\simeq ~ (1-\beta)^j.
\end{equation}

\smallgap Now let us look for lower bounds of $\Delta(t)$.\\
Using the monotonicity of $\psi$ and (\ref{lowerBdTerm_i}), a lower bound follows from (\ref{delta_t_sumFinalBis}), namely
\begin{eqnarray}\label{LB1delta(t)}
\Delta(t) &\ge&
\sum_{j=0}^{i-2}\int_{-\infty}^{~l_{A_j}^j} P\big[B_j~|~R_{A_j}=x\big]\left\{
-\Big[\psi\big(l^{j}_{A_{j+1}}-x~,~t-A_j\big)-\psi\big(l^{j+1}_{A_{j+2}}-x~,~t-A_j\big)\Big]\right.\\
&& \left. \quad +~\bar\psi\big(0~,~t-A_{j+1}\big)\Big[\psi\big(l^{j+1}_{A_{j+1}}-x,A_{j+1}-A_j\big)-\psi\big(l^{j+1}_{A_{j+2}}-x,A_{j+1}-A_j\big)\Big]\right\}~dF_{R_{A_j}}(x)\nonumber\\
&&+~ 1_{(i\le k)}~P(B_{i-1})~\min_{x\le l^{i-1}_{A_{i-1}}}\Big( \psi\big(l^{i}_{A_i}-x~,~ t-A_{i-1} \big)-\psi\big(l^{i-1}_{A_i}-x, t-A_{i-1}\big)\Big).\nonumber
\end{eqnarray}
But we can write, via (\ref{def_AiBisApprox}) and the monotonicity of $\psi$,
\begin{eqnarray*}
\!\!\!\!\!\!&\!\!\!&\!\!\!\int_{-\infty}^{~l_{A_j}^j} P\big[B_j~|~R_{A_j}=x\big]
\Big[\psi\big(l^{j+1}_{A_{j+1}}-x,A_{j+1}-A_j\big)-\psi\big(l^{j+1}_{A_{j+2}}-x,A_{j+1}-A_j\big)\Big]~dF_{R_{A_j}}(x)\\
&\ge&\int_{-\infty}^{~l_{A_j}^j} P\big[B_j~|~R_{A_j}=x\big]
\Big[\bar\psi\big(l^{j+1}_{A_{j+2}}-l_{A_j}^j,A_{j+1}-A_j\big)-\bar\psi\big(l^{j+1}_{A_{j+1}}-x,A_{j+1}-A_j\big)\Big]~dF_{R_{A_j}}(x)\\
&\simeq& P(B_j)~\Big(\bar\psi\big(l^{j+1}_{A_{j+2}}-l^{j}_{A_{j}},A_{j+1}-A_j\big)-(1-\beta)\Big)~=P(B_j)~\Big(\beta-\psi\big(l^{j+1}_{A_{j+2}}-l^{j}_{A_{j}},A_{j+1}-A_j\big)\Big)
\end{eqnarray*}
and using also (\ref{ineqAi}),
\begin{eqnarray*}
&&\int_{-\infty}^{~l_{A_j}^j} \Big(\psi\big(l^{j}_{A_{j+1}}-x~,~t-A_j\big)-~\psi\big(l^{j+1}_{A_{j+2}}-x~,~t-A_j\big)\Big)P\big[B_j~|~R_{A_j}=x\big]~dF_{R_{A_j}}(x)\\
&\le&
\int_{-\infty}^{~l_{A_j}^j} \bar\psi\big(l^{j+1}_{A_{j+1}}-x~,~A_{j+1}-A_j\big)P\big[B_j~|~R_{A_j}=x\big]~dF_{R_{A_j}}(x)~-~\bar\psi\big(l^{j}_{A_{j+1}}-l^{j}_{A_{j}}~,~t-A_j\big)P(B_j)\\
&& -(1-\alpha)(1-\beta)P(B_j)+\max_{x\le l^{j}_{A_{j}}}
\Big(\psi\big(l^{j+1}_{A_{j+1}}-x,A_{j+1}-A_j+d\big)-\psi\big(l^{j+1}_{A_{j+2}}-x,t-A_j\big)\Big)P(B_j)\\
&\simeq&
P(B_j)\Big[1-\beta-\bar\psi\big(l^{j}_{A_{j+1}}-l^{j}_{A_{j}}~,~t-A_j\big) -(1-\alpha)(1-\beta)\\
&&\qquad\quad +\max_{x\le l^{j}_{A_{j}}}
\Big(\psi\big(l^{j+1}_{A_{j+1}}-x,A_{j+1}-A_j+d\big)-\psi\big(l^{j+1}_{A_{j+2}}-x,t-A_j\big)\Big]\\
&=& \!\!\!
\Big[\alpha(1-\beta)-\bar\psi\big(l^{j}_{A_{j+1}}-l^{j}_{A_{j}},t-A_j\big) +\max_{x\le l^{j}_{A_{j}}}\!\!\!
\Big(\psi\big(l^{j+1}_{A_{j+1}}-x,A_{j+1}-A_j+d\big)-\psi\big(l^{j+1}_{A_{j+2}}-x,t-A_j\big)\Big)\Big]P(B_j),
\end{eqnarray*}
hence (\ref{LB1delta(t)}) becomes
\begin{eqnarray}\label{LBdelta(t)defA}
\Delta(t) &\ge& \sum_{j=0}^{i-2}P(B_j)~\Big\{-\alpha+\beta\Big(\alpha+\bar\psi\big(0~,~t-A_{j+1}\big)\Big)-\bar\psi\big(0~,~t-A_{j+1}\big)\psi\big(l^{j+1}_{A_{j+2}}-l^{j}_{A_{j}},A_{j+1}-A_j\big)\nonumber\\
&&+~\bar\psi\big(l^{j}_{A_{j+1}}-l^{j}_{A_{j}},t-A_j\big)
-\max_{x\le l^{j}_{A_{j}}}
\Big(\psi\big(l^{j+1}_{A_{j+1}}-x,A_{j+1}-A_j+d\big)-\psi\big(l^{j+1}_{A_{j+2}}-x,t-A_j\big)\Big)\Big\}\nonumber\\
&&+~ 1_{(i\le k)}~P(B_{i-1})~\min_{x\le l^{i-1}_{A_{i-1}}}\Big( \psi\big(l^{i}_{A_i}-x~,~ t-A_{i-1} \big)~-~\psi\big(l^{i-1}_{A_i}-x, t-A_{i-1}\big)\Big)
\end{eqnarray}
\begin{eqnarray}
\mbox{so}\quad\Delta(t)&\ge& \sum_{j=0}^{i-2}P(B_j)~\Big\{-\alpha+\beta\Big(\alpha+\bar\psi\big(0~,~t-A_{j+1}\big)\Big)-\bar\psi\big(0~,~t-A_{j+1}\big)\psi\big(0~,~A_{j+1}-A_j\big)\nonumber\\
&&\qquad\qquad\qquad +~\bar\psi\big(u_j~,~t-A_j\big)
-\psi\big(u_j~,~A_{j+1}-A_j+d\big)\Big\}\nonumber\\
&&+~ 1_{(i\le k)}~P(B_{i-1})\min_{x\le l^{i-1}_{A_{i-1}}}\Big( \psi\big(l^{i}_{A_i}-x~,~ t-A_{i-1} \big)-\psi\big(l^{i-1}_{A_i}-x, t-A_{i-1}\big)\Big) \label{LBprop}\\
&\ge& \Big\{-\alpha+\beta\Big(\alpha+\bar\psi\big(0,t\big)\Big)
-\bar\psi\big(0,t-A_{i-1}\big)\psi\big(0,\delta^A_i\big)+\bar\psi(0,t)-\psi\big(0,\delta^A_i+d\big)\Big\}\sum_{j=0}^{i-2}P(B_j)\nonumber\\
&&+~ 1_{(i\le k)}~P(B_{i-1})\min_{x\le l^{i-1}_{A_{i-1}}}\Big( \psi\big(l^{i}_{A_i}-x~,~ t-A_{i-1} \big)-\psi\big(l^{i-1}_{A_i}-x, t-A_{i-1}\big)\Big)\nonumber\\
&\simeq& \Big\{-\alpha+\beta\Big(\alpha +\bar\psi\big(0,t\big)\Big)+\bar\psi(0,t)-\psi\big(0,\delta^A_i\big)\bar\psi\big(0,t-A_{i-1}\big)-\psi\big(0,\delta^A_i+d\big)\Big\}~\frac{1-(1-\beta)^{i-1}}{\beta}\nonumber\\
&&+~ 1_{(i\le k)}~(1-\beta)^{i-1}\min_{x\le l^{i-1}_{A_{i-1}}}\Big( \psi\big(l^{i}_{A_i}-x~,~ t-A_{i-1} \big)-\psi\big(l^{i-1}_{A_i}-x, t-A_{i-1}\big)\Big) \label{48}
\end{eqnarray}
\begin{equation}\label{deltaiA}
\mbox{where}\quad \delta^A_i:=\max_{0\le j\le i-2} (A_{j+1}-A_j)
\end{equation}
using (\ref{PBiApprox}) for this last approximation.

\smallgap Now let us look for possible upper bounds of $\Delta(t)$.\\
The integral in $x$ appearing in the second sum of (\ref{delta_t_sumFinalBis}) can be bounded by
\begin{eqnarray*}
&&\int_{-\infty}^{l_{A_j}^j}\Big(\psi\big(l^{j+1}_{A_{j+1}}-x,A_{j+1}-A_{j}\big)-\psi\big(l^{j+1}_{A_{j+2}}-x,A_{j+1}-A_{j}\big)\Big)P\big[B_j~|~R_{A_j}=x\big]~dF_{R_{A_j}}(x)\\
&&=~\int_{-\infty}^{l_{A_j}^j}\Big(\psi\big(l^{j+1}_{A_{j+1}}-x,A_{j+1}-A_{j}\big)-\psi\big(l^{j+1}_{A_{j+1}}-x,A_{j+1}-A_{j}+d\big)\Big)P\big[B_j~|~R_{A_j}=x\big]~dF_{R_{A_j}}(x)\\
&&~+\int_{-\infty}^{l_{A_j}^j}\Big(\psi\big(l^{j+1}_{A_{j+1}}-x,A_{j+1}-A_{j}+d\big)-\psi\big(l^{j+1}_{A_{j+2}}-x,A_{j+1}-A_{j}\big)\Big)P\big[B_j~|~R_{A_j}=x\big]~dF_{R_{A_j}}(x)
\end{eqnarray*}
which, combined with (\ref{delta_t_sumFinalBis}), (\ref{def_AiBis}) and (\ref{upperBdTerm_i}), provides
\begin{eqnarray}
\Delta(t) &\le& \!\!\!-\sum_{j=0}^{i-2}\int_{-\infty}^{~l_{A_j}^j}
\left(\psi\big(l^{j}_{A_{j+1}}-x,t-A_j\big)-\psi\big(l^{j+1}_{A_{j+2}}-x,t-A_j\big)\right)P\big[B_j~|~R_{A_j}=x\big]~dF_{R_{A_j}}(x) \label{BS1deltat}\\
&&\!\!\!+\sum_{j=0}^{i-2}\int_{-\infty}^{l_{A_j}^j}\!\!\!\Big(\psi\big(l^{j+1}_{A_{j+1}}-x,A_{j+1}-A_{j}+d\big)-\psi\big(l^{j+1}_{A_{j+2}}-x,A_{j+1}-A_{j}\big)\Big)P\big[B_j~|~R_{A_j}=x\big]~dF_{R_{A_j}}(x)\nonumber\\
&&\!\!\!-~(1-\alpha)(1-\beta)\sum_{j=0}^{i-2}P(B_j)  +~ 1_{(i\le k)}P(B_{i-1})\max_{x\le l^{i-1}_{A_{i-1}}}\Big( \psi\big(l^{i}_{A_i}-x~,~ t-A_{i-1} \big)-\psi\big(l^{i-1}_{A_i}-x, t-A_{i-1}\big)\Big)\nonumber
\end{eqnarray}
\begin{eqnarray}
\qquad &\le& -~(1-\alpha)~(1-\beta)\sum_{j=0}^{i-2}P(B_j)\label{BS1deltatProp}\\
&&-\sum_{j=0}^{i-2}P(B_j)\min_{x\le l_{A_j}^j}
\Big(\psi\big(l^{j}_{A_{j+1}}-x,t-A_j\big)-\psi\big(l^{j+1}_{A_{j+2}}-x,t-A_j\big)\Big)\nonumber\\
&&+\sum_{j=0}^{i-2}P(B_j)\max_{x\le l_{A_j}^j}\Big(\psi\big(l^{j+1}_{A_{j+1}}-x,A_{j+1}-A_{j}+d\big)-\psi\big(l^{j+1}_{A_{j+2}}-x,A_{j+1}-A_{j}\big)\Big)\nonumber\\
&& +~ 1_{(i\le k)}~P(B_{i-1})\max_{x\le l^{i-1}_{A_{i-1}}}\Big( \psi\big(l^{i}_{A_i}-x~,~ t-A_{i-1} \big)~-~\psi\big(l^{i-1}_{A_i}-x, t-A_{i-1}\big)\Big)\nonumber\\
&\le & \Big\{-~(1-\alpha)~\big(1-\beta)-\min_{0\le j\le i-2}\min_{x\le l_{A_j}^j}
\Big(\psi\big(l^{j}_{A_{j+1}}-x,t-A_j\big)-\psi\big(l^{j+1}_{A_{j+2}}-x,t-A_j\big)\Big) \nonumber\\
&&+\max_{0\le j\le i-2}\max_{x\le l_{A_j}^j}\Big(\psi\big(l^{j+1}_{A_{j+1}}-x,A_{j+1}-A_{j}+d\big)-\psi\big(l^{j+1}_{A_{j+2}}-x,A_{j+1}-A_{j}\big)\Big)\Big\}\sum_{j=0}^{i-2}P(B_j)\nonumber\\
&& +~ 1_{(i\le k)}~P(B_{i-1})\max_{x\le l^{i-1}_{A_{i-1}}}\Big( \psi\big(l^{i}_{A_i}-x~,~ t-A_{i-1} \big)~-~\psi\big(l^{i-1}_{A_i}-x, t-A_{i-1}\big)\Big)\nonumber\\
&\simeq& \Big\{-~(1-\alpha)~\big(1-\beta)-\min_{0\le j\le i-2}\min_{x\le l_{A_j}^j}\Big(\psi\big(l^{j}_{A_{j+1}}-x,t-A_j\big)-\psi\big(l^{j+1}_{A_{j+2}}-x,t-A_j\big)\Big) \nonumber\\
&&+\max_{0\le j\le i-2}\max_{x\le l_{A_j}^j}\Big(\psi\big(l^{j+1}_{A_{j+1}}-x,A_{j+1}-A_{j}+d\big)-\psi\big(l^{j+1}_{A_{j+2}}-x,A_{j+1}-A_{j}\big)\Big)\Big\}\frac{1-(1-\beta)^{i-1}}{\beta}\nonumber\\
&& +~ 1_{(i\le k)}~(1-\beta)^{i-1}\max_{x\le l^{i-1}_{A_{i-1}}}\Big( \psi\big(l^{i}_{A_i}-x~,~ t-A_{i-1} \big)~-~\psi\big(l^{i-1}_{A_i}-x, t-A_{i-1}\big)\Big)\label{UBdeltatmodif}\\
&\le&
\Big(\beta+\alpha(1-\beta)\Big)~\frac{1-(1-\beta)^{i-1}}{\beta}+ \label{UBdeltatmodifBis}\\
&& 1_{(i\le k)}(1-\beta)^{i-1}\!\!\!\max_{x\le l^{i-1}_{A_{i-1}}}\Big( \psi\big(l^{i}_{A_i}-x,t-A_{i-1} \big)-\psi\big(l^{i-1}_{A_i}-x, t-A_{i-1}\big)\Big).\nonumber
\end{eqnarray}
Note that, for $2\le i\le k<+\infty$, this upper bound is equivalent, as $\beta\to 0$ such that $\beta (i-1)<1$, to
$$
\alpha(i-1)+(1-\big(i-1)\beta\big)\max_{x\le l^{i-1}_{A_{i-1}}}\Big( \psi\big(l^{i}_{A_i}-x,t-A_{i-1} \big)-\psi\big(l^{i-1}_{A_i}-x, t-A_{i-1}\big)\Big),
$$ that has to be compared with $1$ (since $\Delta(t)\le 1$).

\smallgap
An alternative way consists in introducing the non negative parameter function of $t$:
\begin{equation}\label{gammat}
\gamma(t):=\min_{0\le j\le i-2}\min_{x\le l^j_{A_j}}\frac{\psi\big(l^{j+1}_{A_{j+2}}-x,A_{j+1}-A_{j}\big)+\psi\big(l^{j}_{A_{j+1}}-x,t-A_{j}\big)}{\psi\big(l^{j+1}_{A_{j+1}}-x,A_{j+1}-A_{j}\big)}\ge 0
\end{equation}
such that for all $x\le l_{A_j~}^j$,
\begin{equation}\label{gammatineq}
\Big[\psi\big(l^{j+1}_{A_{j+2}}-x,A_{j+1}-A_{j}\big)+\psi\big(l^{j}_{A_{j+1}}-x,t-A_{j}\big)\Big]\ge \gamma(t)~\psi\big(l^{j+1}_{A_{j+1}}-x,A_{j+1}-A_{j}\big)
\end{equation}
that might provide a smaller upper bound of $\Delta(t)$ whenever $\gamma(t)> 1$.\\
Indeed, we deduce from (\ref{BS1deltat}) and (\ref{gammatineq}) that
\begin{eqnarray}
\Delta(t)
&\le & 1_{(i\le k)}~P(B_{i-1})\max_{x\le l^{i-1}_{A_{i-1}}}\Big( \psi\big(l^{i}_{A_i}-x~,~ t-A_{i-1} \big)~-~\psi\big(l^{i-1}_{A_i}-x, t-A_{i-1}\big)\Big)\nonumber\\
&& -~(1-\alpha)(1-\beta)\sum_{j=0}^{i-2}P(B_j)\nonumber\\
&& \!\!\!+\sum_{j=0}^{i-2}
\Big(\Big[\psi\big(l^{j+1}_{A_{j+2}}-l_{A_j~}^j,t-A_{j}\big)+\psi\big(l^{j+1}_{A_{j+1}}-l_{A_j~}^j,A_{j+1}-A_{j}+d\big)\Big]P(B_j)\nonumber\\
&& -~\gamma(t)\int_{-\infty}^{l_{A_j~}^j}\!\!\!\psi\big(l^{j+1}_{A_{j+1}}-x,A_{j+1}-A_{j}\big)~P\big[B_j~|~R_{A_j}=x\big]~dF_{R_{A_j}}(x)\Big)\nonumber\\
&\le & 1_{(i\le k)}~P(B_{i-1})\max_{x\le l^{i-1}_{A_{i-1}}}\Big( \psi\big(l^{i}_{A_i}-x~,~ t-A_{i-1} \big)~-~\psi\big(l^{i-1}_{A_i}-x, t-A_{i-1}\big)\Big)\nonumber\\
&& -~(1-\alpha+\gamma(t))(1-\beta)\sum_{j=0}^{i-2}P(B_j)\nonumber\\
&& \!\!\!+\sum_{j=0}^{i-2}P(B_j)\Big[\psi\big(l^{j+1}_{A_{j+2}}-l_{A_j~}^j,t-A_{j}\big)+\psi\big(l^{j+1}_{A_{j+1}}-l_{A_j~}^j,A_{j+1}-A_{j}+d\big)\Big]\label{deltatgammaProp}\\
&\le & 1_{(i\le k)}~P(B_{i-1})\max_{x\le l^{i-1}_{A_{i-1}}}\Big( \psi\big(l^{i}_{A_i}-x~,~ t-A_{i-1} \big)~-~\psi\big(l^{i-1}_{A_i}-x, t-A_{i-1}\big)\Big)\nonumber\\
&& +\Big(2-(1-\alpha+\gamma(t))(1-\beta)\Big)\sum_{j=0}^{i-2}P(B_j) \nonumber\\
&\simeq & 1_{(i\le k)}~(1-\beta)^{i-1}\max_{x\le l^{i-1}_{A_{i-1}}}\Big( \psi\big(l^{i}_{A_i}-x~,~ t-A_{i-1} \big)~-~\psi\big(l^{i-1}_{A_i}-x, t-A_{i-1}\big)\Big)\nonumber\\
&&+\Big(1+\alpha-\gamma(t)+\beta(1-\alpha+\gamma(t))\Big)\frac1\beta \big(1-(1-\beta)^{i-1}\big).\label{deltatgamma}
\end{eqnarray}

\noindent We can conclude to the following propositions.
\newpage
\begin{prop}\label{propOurA}
When comparing the finite time ruin probabilities of our two models $M^k$ (with $k$ alarm times) and $M_r$ in order to evaluate the effectiveness of a strategy involving an alarm system, it comes that, for $2 \le i \le k+1 $ and for  $t\in(A_{i-1}; A_i]$,
$\Delta(t)=\Delta_{k,r}(t)~= ~P(T^{M^k} \le t)~-~ P(T^{M_r}\le t)$ has got various possible lower and upper bounds given above from (\ref{LB1delta(t)}) to (\ref{48}) and  from (\ref{BS1deltat}) to (\ref{deltatgamma}), respectively.
In particular, it satisfies (\ref{BS1deltatProp}), (\ref{deltatgammaProp}) and (\ref{LBprop}), $B_j$ being defined in (\ref{defBi}), namely

\begin{eqnarray*}
\Delta(t)&\le& 1_{(i\le k)}~P(B_{i-1})\max_{x\le l^{i-1}_{A_{i-1}}}\Big( \psi\big(l^{i}_{A_i}-x~,~ t-A_{i-1} \big)~-~\psi\big(l^{i-1}_{A_i}-x, t-A_{i-1}\big)\Big)\\
&& -~(1-\alpha)~(1-\beta)~\sum_{j=0}^{i-2}P(B_j)~ + ~ \min\Big\{-(1-\beta)\gamma(t)~\sum_{j=0}^{i-2}P(B_j)~+ \\
&&\sum_{j=0}^{i-2} \Big[\psi\big(l^{j+1}_{A_{j+2}}-l_{A_j~}^j,t-A_{j}\big)+\psi\big(l^{j+1}_{A_{j+1}}-l_{A_j~}^j,A_{j+1}-A_{j}+d\big)\Big]~P(B_j)~;\\
&& \sum_{j=0}^{i-2} \Big[\max_{x\le l_{A_j}^j}\Big(\psi\big(l^{j+1}_{A_{j+1}}-x,A_{j+1}-A_{j}+d\big)-\psi\big(l^{j+1}_{A_{j+2}}-x,A_{j+1}-A_{j}\big)\Big)\\
&&~ \qquad - \min_{x\le l_{A_j}^j}\Big(\psi\big(l^{j}_{A_{j+1}}-x,t-A_j\big)-\psi\big(l^{j+1}_{A_{j+2}}-x,t-A_j\big)\Big)\Big] ~P(B_j)\Big\}
\end{eqnarray*}
where ~$\displaystyle \gamma(t)=\min_{0\le j\le i-2}\min_{x\le l^j_{A_j}}\frac{\psi\big(l^{j+1}_{A_{j+2}}-x,A_{j+1}-A_{j}\big)+\psi\big(l^{j}_{A_{j+1}}-x,t-A_{j}\big)}{\psi\big(l^{j+1}_{A_{j+1}}-x,A_{j+1}-A_{j}\big)}$ ,\\
and
\begin{eqnarray*}
\Delta(t)&\ge& 1_{(i\le k)}~P(B_{i-1})\min_{x\le l^{i-1}_{A_{i-1}}}\Big( \psi\big(l^{i}_{A_i}-x~,~ t-A_{i-1} \big)-\psi\big(l^{i-1}_{A_i}-x, t-A_{i-1}\big)\Big) \\
&& +~ \sum_{j=0}^{i-2}P(B_j)~\Big[-\alpha+\beta\Big(\alpha+\bar\psi\big(0~,~t-A_{j+1}\big)\Big) -\bar\psi\big(0~,~t-A_{j+1}\big)\psi\big(0~,~A_{j+1}-A_j\big)\\
&& +~\bar\psi\big(u_j~,~t-A_j\big) -\psi\big(u_j~,~A_{j+1}-A_j+d\big)\Big].
\end{eqnarray*}
\end{prop}

{\bf Remarks. }
\begin{itemize}
\item[i)] The proposition recalls explicitly some of the bounds which are easy to compute numerically, even though these are not the sharpest ones proposed. 
The sharper upper bound could be either of the two explicit expressions, depending on time horizon $t$ considered.\\
Note that several alternative bounds have been provided from (\ref{LB1delta(t)}) to (\ref{deltatgamma}); in practice one has to choose the sharpest one, depending on the feasibility of the computations involved, as applicable with the available information. Of course these bounds are useful only when they are strictly less than 1 in absolute value, since $-1\le \Delta (t)\le 1$. We pay greater emphasis on the upper bounds, as these provide indications when the alarm system could be competitive.
\item[ii)] The approximation $\displaystyle P(B_j)\sim (1-\beta)^j$ (see (\ref{PBiApprox})) can be used to simplify computations (e.g. whenever there are many alarms), and leads respectively to the bounds (\ref{48}) and (\ref{deltatgamma}).
\item[iii)] The bounds given in Proposition\ref{propOurA} still hold when considering $t\le A_1$, {\it i.e.} the case $i=1$, reducing to the last term conditioning by $i\le k$.
\item[iv)] An alternative way to evaluate $\Delta(t)$ is to use a direct method whenever $t> A_k$ (main case of interest since the risk of ruin before $A_k$ is small enough for the model $M^k$, to define $k$ alarms). 
We have $\displaystyle P[T^{M_r}>t]=\bar\psi\Big(\sum_{n=0}^k u_ne^{-r{A_n}},t\Big)$ and, using (\ref{psiAi|Bi}),
\begin{eqnarray*}
P[T^{M^k}>t]
&=& P(B_k)\int_{-\infty}^{~l_{A_{k+1}}^{k+1}} \bar\psi\big(l_{A_{k+1}}^{k+1}-x~,~t-A_k\big)P\big[B_k~|~R_{A_k}=x\big]~dF_{R_{A_k}}(x)
\end{eqnarray*}
that satisfies
$$
P^2(B_k)\bar\psi\big(0~,~t-A_k\big)\le ~P[T^{M^k}>t]~\le ~P^2(B_k)
$$
hence
\begin{equation*}
\bar\psi\Big(\sum_{n=0}^k u_ne^{-r{A_n}},t\Big)-P^2(B_k) ~\le ~\Delta(t)~\le ~\bar\psi\Big(\sum_{n=0}^k u_ne^{-r{A_n}},t\Big)-P^2(B_k)\bar\psi\big(0~,~t-A_k\big),
\end{equation*}
where one might use once again (\ref{PBiApprox}) (see \cite{wpDK} for more detail) to get approximations of the bounds.
\end{itemize}

\begin{cor}\label{corRoughBds}
For $1< i\le k<+\infty$ and for $\beta\sim 0$ such that $\beta (i-1)<1$, the bounds proposed for $\Delta(t)$ can be approximated as
\begin{eqnarray*}
\Delta(t) &\le & (i-1)\Big(\alpha+ \min\big\{0~;~1-\gamma(t)+\beta(1+\gamma(t)-\alpha)\big\}\Big)+\\
&&\qquad\quad \big(1-\beta(i-1)\big)\max_{x\le l^{i-1}_{A_{i-1}}}\Big( \psi\big(l^{i}_{A_i}-x,t-A_{i-1} \big)-\psi\big(l^{i-1}_{A_i}-x, t-A_{i-1}\big)\Big)\\
\mbox{and}\quad\Delta(t)&\ge& (i-1)\Big(-\alpha+\bar\psi(0,t)-\psi\big(0,\delta^A_i\big)\bar\psi\big(0,t-A_{i-1}\big)-\psi\big(0,\delta^A_i+d\big)\Big)+\\
&&\qquad\qquad\quad \min_{x\le l^{i-1}_{A_{i-1}}}\Big( \psi\big(l^{i}_{A_i}-x,t-A_{i-1} \big)-\psi\big(l^{i-1}_{A_i}-x, t-A_{i-1}\big)\Big)\\
\mbox{with}&\delta^A_i & \mbox{defined in}~(\ref{deltaiA}).
\end{eqnarray*}
\end{cor}
Note that this last upper bound decreases whenever $\gamma(t)$ increases. In particular, if $\gamma(t)$ is larger than 1, it becomes closer to 0 and might become negative meaning that the probability of survival for an alarm system might become higher than the one for a system without alarms.

\smallgap {\it Revisit Example 2.}\\
Let us revisit Example 2 to obtain a numerical evaluation of those bounds.\\
Choosing for instance $\beta = 0.225$, $\alpha = 0.45$, $d= 1.0$ and initial capital $u_0=50$ with 10\% of it being added at each alarm time, we obtain exactly $k=4$ alarms occurring respectively at $A_1= 0.29$, $A_2= 0.58$, $A_3=0.9$ and $A_4=1.28$. Consider also the rate of interest to be e.g. $r=10\%$. \\
Evaluating the bounds for $\Delta(t)$ at different times $t_i$,  while choosing $A_{i} < t_i < A_{i+1}$, for $i=1,\cdots,4$ (with $A_5=\infty$),
we observe that the upper bound of $\Delta(t)$, as given explicitly in Proposition~\ref{propOurA}, is between 0.30 and 0.48 (see \cite{wpDK} for more details).\\
It is encouraging to observe that the upper bound, even though not the sharpest one, is much smaller than 1 providing indication that the alarm system is competitive in broad generality. This is reinforced through exact numerical evaluation in specific instances, narrated in Section~\ref{compare:numerical}.

\subsubsection{Application in the case of ultimate ruin probability}

This case can be directly deduced from the previous propositions taking $t\to\infty$ which implies to consider also $i=k+1$.\\
All the bounds given from (\ref{LB1delta(t)}) to (\ref{48}) and  from (\ref{BS1deltat}) to (\ref{deltatgamma}) can be rewritten under this hypothesis, in particular we have
\begin{prop}\label{diffDelta}
The difference of ultimate ruin probabilities $\Delta= P[T^{M_r}=\infty]-P[T^{M^k}=\infty]$ admits the following bounds
\begin{eqnarray*}
\Delta&\le&  -~(1-\alpha)~(1-\beta)~\sum_{j=0}^{k-1}P(B_j)~ + ~ \min\Big\{-\gamma(1-\beta)~\sum_{j=0}^{k-1}P(B_j)~+ \\
&&\sum_{j=0}^{k-1} \Big[\psi\big(l^{j+1}_{A_{j+2}}-l_{A_j~}^j\big)+\psi\big(l^{j+1}_{A_{j+1}}-l_{A_j~}^j,A_{j+1}-A_{j}+d\big)\Big]~P(B_j)~;\\
&& \sum_{j=0}^{k-1} \Big[\max_{x\le l_{A_j}^j}\Big(\psi\big(l^{j+1}_{A_{j+1}}-x,A_{j+1}-A_{j}+d\big)-\psi\big(l^{j+1}_{A_{j+2}}-x,A_{j+1}-A_{j}\big)\Big)\\
&&~ \qquad - \min_{x\le l_{A_j}^j}\Big(\psi\big(l^{j}_{A_{j+1}}-x\big)-\psi\big(l^{j+1}_{A_{j+2}}-x\big)\Big)\Big] ~P(B_j)\Big\}
\end{eqnarray*}
where ~ $\displaystyle \gamma=\min_{0\le j\le k-1}\min_{x\le l^j_{A_j}}\frac{\psi\big(l^{j+1}_{A_{j+2}}-x,A_{j+1}-A_{j}\big)+\psi\big(l^{j}_{A_{j+1}}-x\big)}{\psi\big(l^{j+1}_{A_{j+1}}-x,A_{j+1}-A_{j}\big)}$,\\
and
\begin{eqnarray*}
\Delta&\ge& \sum_{j=0}^{k-1}P(B_j)\Big[-\alpha+\beta\Big(\alpha+\bar\psi(0)\Big) -\bar\psi(0)\psi\big(0,A_{j+1}-A_j\big) +\bar\psi(u_j) -\psi\big(u_j,A_{j+1}-A_j+d\big)\Big].
\end{eqnarray*}
\end{prop}

\begin{remark}
Note that if the NPC is violated ($\theta\ge 0$), as in our numerical examples, then  $\bar\psi(u)= 0$, $\forall u$, which implies that $\Delta=0$.\\
Moreover, as for Proposition\ref{propOurA}, the approximation $\displaystyle P(B_j)\sim (1-\beta)^j$ can be used to simplify computations.
\end{remark}

\begin{cor} The following approximations can be deduced.
\begin{itemize}
\item As $\beta\to 0$ and for finite $k$ with $k\beta<1$, we have
\begin{eqnarray*}
\Delta&\le& k\Big(\alpha+ \min\big\{0~;~1-\gamma+(1-\alpha+\gamma)\beta\big\}\Big)\\
\mbox{and}\quad \Delta &\ge & k~\Big(-\alpha+\bar\psi(0)\bar\psi\big(0,\delta^A_{k+1}\big)-\psi\big(0,\delta^A_{k+1}+d\big)\Big)\\
\mbox{with}&&\delta^A_{k+1}=\max_{0\le j\le k-1}(A_{j+1}-A_j).
\end{eqnarray*}
\item As $k\to\infty$, we have
\begin{eqnarray*}
\Delta&\le& \frac1\beta~\Big[\max_{j\ge 0}\max_{x\le l_{A_j}^j}
\Big(\psi\big(l^{j+1}_{A_{j+1}}-x,A_{j+1}-A_{j}\big)- \psi\big(l^{j+1}_{A_{j+2}}-x,A_{j+1}-A_{j}\big)\Big)\\
&& \qquad - \min_{j\ge 0}\min_{x\le l_{A_j}^j}
\left(\psi\big(l^{j}_{A_{j+1}}-x\big)-\psi\big(l^{j+1}_{A_{j+2}}-x\big)\right)-(1-\alpha)(1-\beta)\Big]\\
\mbox{and}\quad\Delta &\ge &\alpha+\bar\psi(0)-\frac{\alpha-\bar\psi(0)\bar\psi\big(0,\delta^A_{k+1}\big)+\psi\big(0,\delta^A_{k+1}+d\big)}{\beta}.
\end{eqnarray*}
\end{itemize}
\end{cor}

\begin{remark}
\begin{itemize}
\item[-] We chose to present approximated bounds easy to compute, although rougher than the ones that could also be deduced from Proposition \ref{diffDelta}.
\item[-] Proceeding in a direct way would provide rougher bounds than the ones obtained by the recursive way. Indeed we can write
$$
\Delta=P[T^{M_r}=\infty]-P[T^{M^k}=\infty]=
\bar\psi\left(\sum_{i=0}^k u_ie^{-r{A_i}}\right)-P\big(B_{k+1}\big)
~\simeq ~\bar\psi\left(\sum_{i=0}^k u_ie^{-r{A_i}}\right)-(1-\beta)^{k+1}
$$
using (\ref{PBiApprox}) for the last approximation.\\
Whenever $\beta (k+1)<1$, we have
$\displaystyle \Delta ~\simeq ~\beta (k+1)-\psi\left(\sum_{i=0}^k u_ie^{-r{A_i}}\right)$,\\
whereas, for $k\to\infty$,
$\displaystyle
\Delta ~\simeq ~1-\psi\left(\sum_{i=0}^{\infty} u_ie^{-r{A_i}}\right)$.\\
Note that if we assume that $\displaystyle u=\sum_{i=0}^k u_ie^{-r{A_i}}\to\infty$, then we can consider for the classical model under NPC the Cram\'er's bound $\displaystyle \psi(u)\sim Ke^{-Cu}$, where $C$ is the adjustment coefficient.
\end{itemize}
\end{remark}

\section{Conclusion}

A simple model has been chosen to illustrate the notion of alarm systems and their use to alleviate the initial capital, adding a complementary capital whenever an alarm would ring.
To validate such a strategy with alarm systems, comparisons have been made between a model with alarms and one without, with equivalent total capital, numerically as well as analytically, providing bounds for the difference of ruin probabilities of the two models.

Our approach has the advantage of being simple and based only on the knowledge on ruin times distributions. Hence it is adaptable to more general models using various L\'evy Processes. This would include cases when the claims are dependent and/or possibly changing distribution, or when the inter-claim time spans follow more complex pattern. However, the specific performance of such an alarm system needs to be closely examined.

Through these adaptations, the proposed alarm system may also be useful for reinsurance companies. Additional considerations and variations in the alarm time formulation for the reinsurance context have been sketched in \cite{wpDK} and will be developed in a future work.

As is well known, (re-) insurance institutions are mandated to periodically monitor and adjust capital, according to Solvency guidelines. Such  reality may be easily integrated into our proposed alarm system by considering a piecewise liner accumulation function, as opposed to a linear one. The exact performance of the alarm system for such an adaptation might be worthwhile to consider.

Another direction where the current work will be expanded is in terms of moving from known deterministic models to a framework where data, as periodically received by the company, would be continually updated and fed into the current framework. This would result in formalizing a more realistic and adaptive alarm system which takes into account all data available up to current time. In principle, such an adaptation from fixed distribution for the severity structure may come via any density estimation procedure, either completely empirical based on available data or an estimate in a Bayesian paradigm. The performance of our alarm system for such an adaptation is obviously of great interest, e.g. in the context of regulation or contracts of contingent capital, and is in the scope of our future work.

\smallgap
{\small {\it Acknowledgement.} The work has been carried out during the first author's stay at ESSEC Business School and the second author's stay at the Indian Institute of Management Bangalore; the authors would like to thank both the institutions for hospitality  and facilitating the research undertaken.}

~\vspace{-2.2cm}\\


\begin{thebibliography}{xx}

\bibitem{AT} Albrecher, H. and and Teugels, J.L. (2007).
On excess of loss Reinsurance. \textit{Eurandom technical report} {\bf
17}.

\bibitem{Asmussen} Asmussen, S. (2000). Ruin probabilities. {\em World Scientific}.

\bibitem{BDMKM} Besson, J-L, Dacorogna, M., de Martin, P., Kastenholz, M.\ and Moller, M.\ (2009). How much capital does a reinsurance need?
\textit{The Geneva Papers} {\bf 34} 159-174.

\bibitem{Borchani} Borchani, A. (2008). Statistiques des valeurs extr\^emes dans le cas de lois discr\`etes. \textit{ESSEC working paper}.

\bibitem{wpDK} Das, S. and Kratz, M. (2010). On Devising Various Alarm Systems for Insurance Companies. \textit{ESSEC-IIMB working paper}.

\bibitem{Dickson} Dickson, D.C. (2005). Insurance Risk and Ruin. {\em Cambridge}.

\bibitem{EKM} Embrechts, P., Kl\"uppelberg, C. and Mikosch, T. (2001). Modelling extremal events for Insurance and Finance. {\em Springer}.

\bibitem{GKL} Guillou, A., Kratz, M. and Le Strat, Y. (2010). An Extreme Value Theory approach for the early detection of time clusters with application to the surveillance of Salmonella. \textit{Preprint ArXiv 1003.4466}

\bibitem{GNDR} Guillou, A., Naveau, P., Diebolt J. and Ribereau, P. (2009). Return level bounds for small and moderate sample sizes for discrete and continuous random variables. {\em Test} {\bf 18} 584-604.

\bibitem{IK04} Ignatov, Z.G and Kaishev, V.K. (2004). A finite-time ruin probability formula for continuous claim severities. \textit{J. Appl. Probab.} {\bf 41 } 570--578.

\bibitem{IK06} Ignatov, Z.G and Kaishev, V.K. (2006). On the infinite horizon probability of (non-)ruin for integer valued claims. \textit{J. Appl. Probab.} {\bf 43} 535-551.

\bibitem{IKK01} Ignatov, Z.G, Kaishev, V.K. and Krachunov, R.S. (2001). An improved finite-time ruin probability formula and its {\it Mathematica} implementation. \textit{Insurance: Mathematics and Economics} {\bf 39} 376--389.


\bibitem{KD} Kaishev, V.K. and Dimitrova, S.D. (2006).
Excess of loss reinsurance under joint survival optimality.
\textit{Insurance: Mathematics and Economics} {\bf 39 (3)} 376--389.

\bibitem{KDI} Kaishev, V.K., Dimitrova, S.D. and Ignatov, Z.G. (2007).
Operational Risk and Insurance: a ruin-probabilistic reserving approach.
\textit{Astin} 2007.

\bibitem{Lindgren} Lindgren, G. (1980). Model processes in non-linear prediction, with application to detection and alarm. \textit{Ann. Probab.} {\bf 8} 775--792.

\bibitem{McFE} McNeil, A.J., Frey R. and Embrechts, P. (2005). Quantitative Risk Management. {\em Princeton}.

\bibitem{Mikosch} Mikosch, T. (2004). Non-life Insurance Mathematics. {\em Springer}.

\bibitem{Scotto} Monteiro, M., Pereira, I. and Scotto, M. (2008). Optimal alarm systems for count processes. \textit{Communications in Statistics - Theory and Methods} {\bf 37 (19)} 3054--3076.

\bibitem{Schmidli} Schmidli, H. (2000). Characteristic of ruin probabilities in classical risk models with and without investment, Cox risk models and perturbed risk models. \textit{Memoirs No 15, Dept. Theor. Statist., Aarhus University}.

\bibitem{ZD2O}Zumbach, G., Dacorogna, M., Olsen, J. and Olsen, R. (2000).
Measuring Shocks in Financial Markets. \textit{Journal of Theoretical and Applied Finance}, {\bf 3(3)} 347--355.
\end{thebibliography}
\end{document}